\documentclass[10pt]{article}
\usepackage[utf8]{inputenc}
\usepackage{fancyhdr}
\usepackage{extramarks}
\usepackage{amsmath}
\usepackage{amsthm}
\usepackage{amsfonts}
\usepackage{siunitx} 
\usepackage{tikz}
\usepackage[plain]{algorithm}
\usepackage{algpseudocode}
\usepackage{multirow}
\usepackage{booktabs}
\usepackage{palatino}
\usepackage{graphicx}
\usepackage{subcaption}
\usepackage[colorlinks,linkcolor=black,anchorcolor=black,citecolor=black,urlcolor=blue]{hyperref}
\usepackage{amsmath,bm}
\usepackage{booktabs}
\usepackage{mathtools}
\usepackage{amssymb}
\usepackage{caption}
\usepackage{capt-of}
\usepackage{mciteplus}
\usepackage{cite}
\usepackage{mathrsfs}
\usepackage[title,titletoc,toc]{appendix}
\usepackage{xr}
\usepackage{parskip}
\usepackage{textcomp}
\usepackage[colaction]{multicol}
\usepackage[switch]{lineno}
\usepackage{lipsum}
\usepackage{etoolbox}
\usepackage{longtable}
\usepackage{array}
\usepackage{tablefootnote}

\newcolumntype{C}[1]{>{\centering\arraybackslash}p{#1}}
\usetikzlibrary{automata,positioning}
\usetikzlibrary{automata,positioning}

\newcommand{\tabincell}[2]{\begin{tabular}{@{}#1@{}}#2\end{tabular}}  
\begin{document}

\title{UMAP-assisted $K$-means clustering of large-scale SARS-CoV-2 mutation datasets}
%Dimension-reduced clustering of large-scale SARS-CoV-2 genome datasets
\author{Yuta Hozumi$^1$, Rui Wang$^1$, Changchuan Yin$^2$, and Guo-Wei Wei$^{1,3,4}$\footnote{
		Corresponding author.		E-mail: weig@msu.edu} \\% Author name
	$^1$ Department of Mathematics, \\
	Michigan State University, MI 48824, USA.\\
	$^2$ Department of Mathematics, Statistics, and Computer Science, \\
	University of Illinois at Chicago, Chicago, IL 60607, USA\\
	$^3$ Department of Electrical and Computer Engineering,\\
	Michigan State University, MI 48824, USA. \\
	$^4$ Department of Biochemistry and Molecular Biology,\\
	Michigan State University, MI 48824, USA. \\
}

\date{} % Date for the report

\maketitle

\begin{abstract}
Coronavirus disease 2019 (COVID-19) caused by severe acute respiratory syndrome coronavirus 2 (SARS-CoV-2) has a worldwide devastating effect. 
The understanding of evolution and transmission of SARS-CoV-2 is of paramount importance for the COVID-19 control, combating, and prevention. Due to the rapid growth of both the number of SARS-CoV-2 genome sequences and the number of unique mutations, the phylogenetic analysis of SARS-CoV-2 genome isolates faces an emergent large-data challenge. We introduce a dimension-reduced $k$-means clustering strategy to tackle this challenge.  
We examine the performance and effectiveness of three dimension-reduction algorithms: principal component analysis (PCA), t-distributed stochastic neighbor embedding (t-SNE), and uniform manifold approximation and projection (UMAP). By using four benchmark datasets, we found that UMAP is the best-suited technique due to its stable, reliable, and efficient performance, its ability to improve clustering accuracy, especially for large Jaccard distanced-based datasets, and its superior clustering visualization. The UMAP-assisted $k$-means   clustering  enables us to  shed light on increasingly large datasets from SARS-CoV-2 genome isolates.

\end{abstract}

\pagenumbering{roman}
\begin{verbatim}
\end{verbatim}

{\setcounter{tocdepth}{4} \tableofcontents}
\newpage
 %\clearpage
 %\pagebreak

% {\setcounter{tocdepth}{4} \tableofcontents}

\setcounter{page}{1}
\renewcommand{\thepage}{{\arabic{page}}}
%\linenumbers

\section{Introduction}
Beginning in December 2019, coronavirus disease 2019 (COVID-19) caused by severe acute respiratory syndrome coronavirus 2 (SARS-CoV-2) has become one of the most deadly global pandemic in history.  The COVID-19 infections in the US and other nations are still spiking. As of October 30, 2020, the World Health Organization (WHO) has reported  44,888,869 confirmed cases of COVID-19 and 1,178,145 confirmed deaths. The virus has spread to Africa, Americas, Eastern Mediterranean, Europe, South-East Asia and Western Pacific \cite{weeklyReport}. To prevent further damage to our livelihood, we must control its spread through testing, social distancing, tracking the spread, and developing effective vaccines, drugs, diagnostics, and treatments.

SARS-CoV-2 is a positive-sense single-strand RNA virus that belongs to the Nidovirales order, coronaviridae family and betacoronavirus genus \cite{of2020species}. To effectively track the virus, testing patients with suspected exposure to COVID-19 and sequencing the strand via PCR (polymerase chain reaction) are important. From sequencing, we can analyze patterns in mutation and predict transmission pathways. Without understanding such pathways, current efforts to find effective medicines and vaccines could become futile because  mutations may change viral genome or lead to resistance. As of October 30, 2020, there are 89627 available sequences with  23763 unique single nucleotide polymorphisms (SNPs) with respect to the first SARS-CoV-2 sequent collected in December 2019 \cite{wu2020new} according to our mutation tracker \url{https://users.math.msu.edu/users/weig/SARS-CoV-2_Mutation_Tracker.html}.

A popular method for understanding mutational trends is to perform phylogenetic analysis, where one clusters mutations to find evolution patterns and transmission pathways. Phylogenetic analysis has been done on the Nidovirales family \cite{alam2020functional, gong2020sars, forster2020phylogenetic, li2020evolutionary, alam2020functional, kasibhatla2020understanding} to understand genetic evolutionary pathways, protein level changes \cite{wang2020decoding, wang2020characterizing, chen2020mutations, kasibhatla2020understanding}, large scale variants \cite{wang2020characterizing, wang2020decoding, wang2020decoding0,worobey2020emergence} and  global trends \cite{ bai2020comprehensive, toyoshima2020sars, van2020emergence}. Commonly used  techniques for phylogenetic analysis include tree based methods \cite{page2012space} and $K$-means clustering.  Both methods belong to unsupervised machine learning techniques, where ground truth is unavailable.  These approaches provide valuable information for exploratory research. A main issue with phylogenetic tree analysis is that as the number of samples increase, its computation becomes unpractical, making it unsuitable for large genome datasets. In contrast, $K$-means scales well with sample size increase, but does not perform well when the sample size is too small. Jaccard distance is commonly used to compare genome sequences \cite{zhou2008approach} because it offers a phylogenetic or topological difference between samples. However, the tradeoff to the Jaccard distance  is that its feature dimension is the same as its  number of samples, suggesting that for a large sample size, the number of features is also large. Since $K$-means clustering relies on computing the distance between the center of the cluster and each sample, having a large feature space can result in expensive computation, large memory requirement, and poor clustering performance. This become a significant problem as the number of SARS-CoV-2 genome isolates from patients has reached 150,000 at this point. There is a pressing need for efficient clustering methods for SARS-CoV-2 genome sequences.   

One technique to address this challenge is to perform dimensional reduction on the $K$-means input dataset so that the task becomes manageable. Commonly used dimension reduction algorithms focus on two aspects: 1) the pairwise distance structure of all the data samples and 2) preservation of the local distances over the global distance. Techniques such as principal component analysis (PCA) \cite{jolliffe2016principal}, Sammon mapping \cite{sammon1969nonlinear}, and multidimensional scaling (MDS) \cite{cox2008multidimensional} aim to preserve the pairwise distance structure of the dataset. In contrast, the t-distributed stochastic neighbor embedding (t-SNE) \cite{linderman2019fast, maaten2008visualizing}, uniform manifold approximation and projection (UMAP) \cite{mcinnes2018umap,becht2019dimensionality}, Laplacian eigenmaps \cite{belkin2001laplacian}, and LargeVis \cite{tang2016visualizing}  focus on the preservation of local distances. Among them, PCA, t-SNE, and UMAP are the most frequently used algorithms in the applications of cell biology, bioinformatics, and visualization \cite{becht2019dimensionality}. 
 
PCA is a popular method used in exploratory studies, aiming to find the directions of the maximum variance in high-dimensional data and projecting them onto a new subspace to obtain low-dimensional feature spaces while preserving most of the variance. The principal components of the new subspace can be interpreted as the directions of the maximum variance, which makes the new feature axes orthogonal to each other. Although PCA is able to cover the maximum variance among features, it may lose some information if one chooses an inappropriate number of principal components. As a linear algorithm, PCA performs poorly on the features with nonlinear relationship. Therefore, in order to present high-dimensional data on low dimensional and nonlinear manifold, some nonlinear dimensional reduction algorithms such as t-SNE and UMAP are employed. T-SNE is a nonlinear method that can preserve the local and global structures of data. There are two main steps in t-SNE.
First, it finds a probability distribution of the high dimensional dataset, where similar data points are given higher probability. Second, it finds a similar probability distribution in the lower dimension space, and the difference between the two distributions is minimized. However, t-SNE computes pairwise conditional probabilities for each pair of samples and involves hyperparameters that are not always easy to tune, which makes it computationally complex. UMAP is a novel manifold learning technique that also captures a nonlinear structure, which is competitive with t-SNE for visualization quality and maintains more of the global structure with superior run-time performance \cite{mcinnes2018umap}. UMAP is built upon the mathematical work of Belkin and Niyogi on Laplacian eigenmaps, aiming to address the importance of uniform data distributions on manifolds via Riemannian geometry and the metric realization of fuzzy simplicial sets by David Spivak \cite{spivak2012metric}. Similar to t-SNE, UMAP can optimize the embedded low-dimensional representation  with respect to fuzzy set cross-entropy loss function by using stochastic gradient descent. The embedding is found by finding a low-dimensional projection of the data that closely matches the fuzzy topological structure of the original space. The error between two topological spaces will be minimized by optimizing the spectral layout of data in a low dimensional space.

In this work, we explore efficient computational methods for the SARS-CoV-2 phylogenetic analysis of large volumn of SARS-CoV-2 genome sequences. Specifically, we are interested in developing a dimension-reduction assisted clustering method. To this end, we compare the effectiveness and accuracy of PCA, t-SNE and UMAP for dimension reduction in association with the $K$-means clustering. To quantitatively evaluate the performance, we recast supervised classification problems with labels into a $K$-means clustering problems so that the accuracy of $K$-means clustering can be evaluated. As a result, the accuracy and performance of PCA, t-SNE and UMAP-assisted $K$-means clustering can be compared. By choosing the different dimensional reduction ratios, we examine the performance of these methods in $K$-means settings on four standard datasets. We found that UMAP is the most efficient, robust, reliable, and accurate algorithm. Based on this finding, we applied the UMAP-assisted $K$-means technique to large scale SARS-CoV-2 datasets generated from a Jaccard distance representation and a SNP position-based representation to further analyze its effectiveness, both in terms of speed and scalability. Our results are compared with those in the literature  \cite{wang2020decoding} to shed new light on SARS-CoV-2 phylogenetics.

\section{Methods}

\subsection{Sequence and alignment}
The SARS-CoV-2 sequences were obtained from GISAID databank (\url{www.gisaid.com}). Only complete genome sequences with collection date, high coverage, and without 'NNNNNN' in the sequences were considered. Each sequence was aligned to the reference sequence \cite{wu2020new} using a multiple sequence alignment (MSA) package Clustal Omega \cite{sievers2011fast}. A total of 23763 complete SARS-CoV-2 sequences are analyzed in this work.

\subsection{SNP position based features}
Let $N$ be the number of SNP profiles with respect to the SARS-CoV-2 reference genome sequence, and let $M$ be the number of unique mutation sites. Denote $V_i$ as the position based feature of the $i$th SNP profile.
\begin{equation}
V_i = [v_i^1, v_i^2, ..., v_i^M], \quad i = 1,2,..., N
\end{equation}
is a $1\times M$ vector. Here
\begin{equation}
v_i^j = \begin{cases}
1, & \text{mutation site} \\
0, & \text{otherwise}. \\
\end{cases}
\end{equation}
We compile this into an $N\times M$ position based feature,
\begin{equation}
S(i,j) = v_i^j
\end{equation}
where each row represents a sample. Note that $S(i,j)$ is a binary representation of the position and is sparse.

\subsection{Jaccard based representation}\label{sec:Jaccard}

The Jaccard distance measures the dissimilarity between two sets. It is widely used in the phylogenetic studies of SNP profiles. In this work, we utilize Jaccard distance to compare SNP profiles of SARS-CoV-2 genome isolates.

Let $A$ and $B$ be two sets. Consider the Jaccard index between $A$ and $B$, denoted $J(A,B)$,  as the cardinality of the intersection divided by the cardinality of the union
\begin{equation}
	J(A,B) = \frac{\big| A\cap B\big|}{\big|A \cup B \big|} = \frac{\big| A\cap B\big|}{\big|A\big| + \big|B\big| - \big|A \cap B \big|}. 
\end{equation}
The Jaccard distance between the two sets is defined by subtracting the Jaccard index from 1:
\begin{equation}
	d_J(A,B) = 1- J(A,B) = \frac{\big|A \cup B \big| - \big|A \cap B \big|}{\big|A \cup B \big|}
\end{equation}

We assume there are $N$ SNP profiles or genome isolates that have been aligned to the reference SARS-CoV-2 genome. Let $S_i$, $i=1,...,N$, be the set with the position of the mutation of the $i$th sample. The Jaccard distance between two sets $S_i$ and $S_j$ is given by $d_J(S_i, S_j)$. Taking the pairwise distance between all the samples, we can construct the Jaccard based representation, resulting in an $N\times N$ distance  matrix $D$
\begin{equation}
	D(i,j) = d_J(S_i, S_j)
\end{equation}
This distance defines  a metric over the collections of all finite sets \cite{levandowsky1971distance}.

%
%\subsection{$k$-nearest neighbor ($k$-NN)}
	%$k$-nearest neighbor ($k$-NN) is a supervised machine learning technique used for both classification and regression. In this section, the $k$-NN for classification is discussed. 
	%
	%The goal of $k$-NN is to classify unknown data point using $k$ nearest known data point. Let $X \in \mathbb{R}^{N,M}$ be the dataset, where $N$ is the number of data, and $M$ is the number of features. Let $x_i$ $i = 1,2,...,N$ represent the $i$th data, and $x_i \in \mathbb{R}^{N}$. Let $k$ be the hyperparameter. Define $\{y_j\}_{j=1^n}$, $n < N$ be the data with ground truth. For each $i$, computed the distance between the $x_i$ and $y_j$ for all $j$.
	%\begin{equation}
	%d(x_i,y_j) = \|x_i - y_j\|
	%\end{equation}
	%where $\|\cdot\|$ defines a distance metric such as Euclidean distance, Manhattan distance, Minkowski distance, Chebyshev distance, Natural log distance, Generalized exponential distance, Generalized Lorentzian distance, Camberra distance, Quadratic distance, and Mahalanobis distance. The $k$ smallest distance, and the class with the most number of closest points is assigned to $x_i$. This process is repeated for all $i$.
	%

\subsection{$K$-means clustering}
$K$-means clustering is one of the most popular unsupervised learning methods in machine learning, where it aims to cluster or partition a data $\{x_1, ..., x_N\}$, $x_i \in \mathbb{R}^M$ into $k$ clusters, $\{C_1, ..., C_k\}$, $k \le N$.

$K$-means clustering begins with selecting $k$ points as $k$ cluster centers, or centroids. Then, each point in the dataset is assigned to the nearest centroid. The centroids are then updated by minimizing the within-cluster sum of squares (WCSS), which is defined as 
\begin{equation}
\sum_{i=1}^k \sum_{x_i \in C_k} \|x_i - \mu_k\|_2^2.
\end{equation}
Here, $\|\cdot\|_2$ denotes the $l_2$ norm and $\mu_k$ is the average of the data point in cluster $k$
\begin{equation}
\displaystyle \mu_k = \frac{1}{|C_k|} \sum_{x_i \in C_k} x_i.
\end{equation}

This method, however, only finds the optimal centroid, given a fixed number of clusters $k$. In applications, we are interested in finding the optimal number of clusters as well. In order to obtain the best $k$ clusters, elbow method was used. The optimal number of clusters can be determined via the elbow method by plotting the WCSS against the number of clusters, and choosing the inflection point position as the optimal number of clusters.

\subsection{Principal component analysis}
Principal component analysis (PCA) is one the most commonly used  dimensional reduction techniques for the exploratory analysis of high-dimensional data \cite{jolliffe2016principal}. Unlike other methods, there is no need for any assumptions in the data. Therefore, it is a useful method for new data, such as  SARS-CoV-2 SNPs data. PCA is conducted by obtaining one component or vector at a time. The first component, termed the principal component, is the direction that maximizes the variance. The subsequent components are orthogonal to earlier ones.

Let $\{x_i\}_{i=1}^N$ be the input dataset, with $N$ being the number of samples or data points. For each $x_i$, let $x_i \in \mathbb{R}^M$, where $M$ is the number of features or data dimension. Then, we can cast the data as a matrix $X \in \mathbb{R}^{N \times M}$. PCA seeks to find a linear combination of the columns of $X$ with maximum variance.
\begin{equation}
\sum_{j=1}^n a_jx_j = Xa,
\end{equation}
where $a_1, a_2, ..., a_n$ are constants. The variance of this linear combination is defined as 
\begin{equation}
{\rm var}(Xa) = a^TSa,
\end{equation}
where $S$ is the covariance matrix for the dataset. Note that we compute the eigenvalue of the covariance matrix. The maximum variance can be computed iteratively using Rayleigh's quotient
\begin{equation}
a_{(1)} = \arg \max_a \frac{a^TX^TXa}{a^Ta}.
\end{equation}
The subsequent components can be computed by maximizing the variance of
\begin{equation}
\hat{X}_k = X - \sum_{j=1}^{k-1} Xa_ja_{j}^T
\end{equation}
where $k$ represents the $k$th principal component. Here, $k-1$ principal components are subtracted from the original matrix $X$. Therefore, the complexity of the method scales linearly with the number of components one seeks to find. In applications, we hope that the first few components give rise to a good PCA representation of the original data matrix $X$.

\subsection{t-SNE}
The t-distributed stochastic neighbor embedding (t-SNE) is a nonlinear dimensional reduction algorithm that is well suited for reducing high dimensional data into the two- or three-dimensional space. There are two main stages in t-SNE. First, it constructs a probability distribution over pairs of data such that a pair of near data points is assigned with a high probability, while a pair of farther away points is given a low probability. Second, t-SNE defines a probability distribution in the embedded space that is similar to that in the original high-dimensional space, and aims to minimize the Kullback-Leibler (KL) divergence between them \cite{linderman2019fast}.

Let $\{x_1, x_2, ..., x_N | x_i \in \mathbb{R}^M \}$ be a high dimensional input dataset. Our goal is to find an optimal  low dimensional representation $\{y_1, ..., y_N | y_i \in \mathbb{R}^k\}$, such that $k << M$.
The first step in t-SNE is to compute the pairwise distribution between $x_i$ and $x_j$, defined as $p_{ij}$. However, we find the conditional probability of $x_j$, given $x_i$:
\begin{equation}\label{pdist} 
p_{j|i} = \frac{\exp(-\|x_i - x_j\|^2/2\sigma_i^2)}{\sum_{m\ne i}\exp(-\|x_i - x_m\|^2/2\sigma_i^2)}, \quad i \ne j,
\end{equation}
setting $p_{i|i} = 0$, and the denominator normalizes the probability. Here, $\sigma_i$ is the predefined hyperparameter called perplexity.  A smaller $\sigma_i$ is used for a denser dataset. Notice that this conditional probability is symmetric when the perplexity is fixed, i.e. $p_{i|j} = p_{j|i}$. Then, define the pairwise probability as 
\begin{equation}
p_{ij} = \frac{p_{j|i} + p_{ij}}{2N}.
\end{equation}

In the second step, we learn a $k$-dimensional embedding $\{y_1, ..., y_N | y_i \in \mathbb{R}^k\}$. To  this end, t-SNE calculates a similar probability distribution $q_{ij}$ defined as
\begin{equation}\label{qdist} 
q_{ij} = \frac{\frac{1}{1+\|y_i - y_j\|^2}}{\sum_{m}\sum_{l\ne m}\frac{1}{1+\|y_m - y_l\|^2}}, \quad i \ne j
\end{equation}
and setting $q_{ii} = 0$. Finally, the low dimensional embedding $\{y_1, ..., y_N | y_i \in \mathbb{R}^k\}$ is found by minimizing the KL-divergence via a standard gradient descent method
\begin{equation}
{\rm KL}(P|Q) = \sum_{i, j} p_{ij}\log\frac{p_{ij}}{q_{ij}},
\end{equation}
where $P$ and $Q$ are the distributions for  $p_{ij}$ and  $q_{ij}$, respectively. Note that the probability distributions in Eqs. (\ref{pdist}) and  (\ref{qdist}) can be replaced by using many other delta sequence kernel of positive type  \cite{wei2000wavelets}. 

\subsection{UMAP}
Uniform manifold approximation and projection (UMAP) is a nonlinear dimensional reduction method, utilizing three assumptions: the data is uniformly distributed on Riemannian manifold, Riemannian metric is locally constant, and the manifold if locally connected. Unlike t-SNE which utilizes probabilistic model, UMAP is a graph-based algorithm. Its essentially idea is to create a predefined $k$-dimensional weighted UMAP graph representation of each of the original high-dimensional data point such that the edge-wise cross-entropy between the weighted graph and the original data is minimized. Finally, the $k$-dimensional eigenvectors of the UMAP graph are used to represent each of the original data point. 
In this section, a computational view of UMAP is presented. For a more theoretical account, the reader is referred to Ref. \cite{mcinnes2018umap}.

Similar to t-SNE, UMAP considers the input data $X = \{x_1, x_2, ..., x_N\},$   $x_i \in \mathbb{R}^M$ and look for  an optimal  low dimensional representation $\{y_1, ..., y_N | y_i \in \mathbb{R}^k\}$, such that $k < M$. The first stage is the construction of  weighted $k$-neighbor graphs. Let  define a metric $d: X\times X \to \mathbb{R}^+$. Let $k << M$ be a hyperparemeter, and compute the $k$-nearest neighbors of each $x_i$ under a given metric $d$. For each $x_i$, let
\begin{equation}
\rho_i = \min\{d(x_i, x_j)| 1 \le j \le k, d(x_i, x_j) > 0\}
\end{equation}
where $\sigma_i$ is defined via
\begin{equation}
\sum_{j=1}^k \exp\left(\frac{-\max(0, d(x_i, x_j) - \rho_i)}{\sigma_i}\right) = \log_2 k.
\end{equation}
One chooses $\rho_i$ to ensure at least one data point is connected to $x_i$ and having edge weight of 1, and set  $\sigma_i$ as a length scale parameter.
One defines a weighted directed graph $\bar{G} = (V, E, \omega)$, where $V$ is the set of vertices (in this case, the data $X$),  $E$ is the set of edges $E = \{ (x_i, x_j)| 1 \le h \le k, 1 \le i \le N\}$, and   $\omega$ is the weight for edges
\begin{equation}
\omega(x_i, x_j) = \exp\left( \frac{-\max(0, d(x_i, x_j) - \rho_i)}{\sigma_i}\right).
\end{equation}
UMAP tries to define an undirected weighted graph $G$ from  directed graph $\bar{G}$ via symmetrization. Let $A$ be the adjacency matrix of the graph $\bar{G}$.   A symmetric matrix can be obtained
\begin{equation}
B = A + A^T - A\otimes A^T,
\end{equation}
where $T$ is the transpose and $\otimes$ denotes the Hadamard product. Then, the undirected weighted Laplacian $G$ (the UMAP graph) is defined by its adjacency matrix $B$. 

In its realization, UMAP evolves an equivalent weighted graph $H$ with a set of points $\{y_i\}_{i=1,\cdots,N}$, utilizing attractive and repulsive forces. The attractive and repulsive forces at coordinate $y_i$ and $y_j$ are given by
\begin{align}
& \frac{-2ab \|y_i - y_j\|_2^{2(b-1)}}{1 + \|y_i - y_j\|_2^2} w(x_i, x_j) (y_i - y_j), ~~{\rm and} \\
& \frac{2b}{(\epsilon + \|y_i - y_j\|_2^2)(1+a\|y_i - y_j\|_2^{2b})} (1-w(x_i, x_j)) (y_i - y_j)
\end{align}
where $a,b$ are hyperparemeters, and $\epsilon$ is taken to be a small value such that the denominator does not become 0. The goal is to find the optimal low-dimensional coordinates $\{y_i\}_{i=1}^{N}$, $y_i \in \mathbb{R}^k$, that minimizes the edge-wise cross entropy with the original data at each point. The evolution of the UMAP graph Laplacian $G$ can be regarded as a discrete approximation of the Laplace-Beltrami operator on a manifold defined by the data \cite{chen2019evolutionary}. Implementation and further detail of UMAP can be found in Ref. \cite{mcinnes2018umap}. 

UMAP may not work well if the data points is non-uniform. If part of the data points have $k$ important neighbors while other part of the data points have $k'>>k$ important neighbors, the $k$-dimensional UMAP will not work efficiently. Currently, there is no algorithm to automatically determine the critic minimal $k_{\rm min}$ for a given dataset. Additionally, weights $w(x_i,x_j)$ and force terms can be replaced by other functions that are easier to evaluate \cite{wei2000wavelets}. The metric $d$ can be selected as Euclidean distance, Manhattan distance, Minkowski distance, and Chebyshev distance, depending on applications. 

\section{Validation}
 $K$-means clustering is one of the unsupervised learning algorithms, suggesting that neither the accuracy nor the root-mean-square error can be calculated to evaluate the performance of the $K$-means clustering explicitly. Additionally, $K$-means clustering can be problematic for high-dimensional large datasets. Dimension-reduced $K$-means clustering is an efficient approach.  
To evaluate its accuracy and performance, we convert supervised classification problems with known salutations into dimension-reduced $K$-means clustering problems. In doing so,  we apply the $K$-means clustering to the classification dataset by setting the number of clusters equals to the number of the real categories. Next, in each cluster, we will take the data with the dominant label as the test for all samples and then calculate the $K$-means clustering accuracy for the whole dataset. %For simplicity, we will call such an implicit accuracy as the "$K$-means accuracy".
 
\subsection{Validation data}\label{sec:Validation}
In this work, we will consider the following classification datasets to test the performance of the clustering methods: Coil 20, Facebook large page-page network, MNIST, and Jaccard distanced-based MNIST. 
\begin{itemize}
	\item \textbf{Coil 20}: Coil 20 \cite{Coil20} is a dataset with 1440 gray scale images, consisting of 20 different objects, each with 72 orientation. Each image is of size $128\times 128$, which was treated as a 16384 dimensional vector for dimensional reduction
	
	\item \textbf{Facebook Network}: Facebook large page-page network \cite{rozemberczki2019multiscale} is a page-page webgraph of verified Facebook sites. Each node represents a facebook page, and the links are the mutual links between sites. This is a binary dataset with 22,470 nodes; hence the sample size and feature size are both 22,470. Jaccard distance was computed between each nodes for the feature space.
	
	\item \textbf{MNIST}: MNIST \cite{lecun1998gradient} is a hand written digit dataset. Each image is a grey scale of size $28\times 28$, which was treated as a 784 dimensional vector for the feature space, each with a integer value in [0, 255]. Standard normalization was used before performing dimensional reduction. There are 70,000 sample, with 10 different labels.
	
	\item \textbf{Jaccard distanced-based MNIST}: The above dataset was converted to a Jaccard distance-based dataset. This is to simulate position based mutational dataset, where 1 indicates a mutation in a particular position. Jaccard distance was used to construct the feature space, hence for each sample, the feature size is 70,000. This dataset can be viewed as an additional validation on our Jaccard distance representation. 
	
\end{itemize}

\subsection{Validation results}
In the present work, we implement three popular dimensional reduction methods, PCA, UMAP, and t-SNE, for the dimension reduction and compare their performance in $K$-means clustering. For a uniform comparison, we reduce the dimensions of the samples by a set of ratios.  The minimum between the number of features and the number of samples was taken as base of the reduction. For the Coil 20 dataset, since the numbers of samples and features were 1440 and 16384, respectively,  dimension-reductions were based on 1440. 
For the Facebook Network,  since the numbers of samples and features were both 
22,470, dimension-reductions were based on 22,470.
For the MNIST dataset,  since the numbers of samples and features were respectively 70,000 and 784, dimension-reductions were based on 784. Finally,
 for the Jaccard distanced-based MNIST dataset, since the numbers of samples and features were both 70,000, dimension-reductions were based on 70,000. 
Note that for the Jaccard distanced-based MNIST data, more aggressive ratios were used because the original feature size is huge, i.e., 70,000. The standard ratios of 2, 4, and 8, etc do not sufficiently reduce the dimension  for effective $K$-means computation. For the purpose of visualization, two-dimensional reduction algorithms are applied to each reduction scheme. 
In order to validate PCA, UMAP, and t-SNE assisted $K$-means clustering, we observed their performance using  labeled datasets. $K$-nearest neighbors ($K$-NN) was used to find the baseline of the reduction, which reveals how much information can be preserved in the feature after applying a dimensional reduction algorithm. For $k$-NN, 10 fold cross-validation was performed.

Notably, $K$-means clustering is an unsupervised learning algorithm, which does not have labels to evaluate the clustering performance explicitly. However, we can assess the $K$-means clustering accuracy via  labeled datasets that has ground truth. In doing so, we choose the number of $K$ as the original number of classes. Then, we can compared the $k$-means clustering results  with the ground truth.  Therefore, the accuracy can reveal the performance of the proposed dimension-reduction-assisted ($k$-means) clustering method. For the classification problem, we assume the training set is $\{(\mathbf{x}_i, y_i) | \mathbf{x}_i\in \mathbb{R}^m, y_i \in\mathbb{Z} \}_{i=1}^{n}$ with the $|\{y_i\}_{i=1}^{n}|=k$. Here $n$, $m$, and $k$ represent the number of samples, the number of features $\{\mathbf{x}_i\}$, and the number of labels $\{y_i\}$, respectively. We set the number of clusters equals to the number of labels $k$. After applying the $K$-means clustering algorithm, we  get $k$ different clusters $\{\mathbf{c}_j\}^{k}_{j=1}$. In each cluster, we define the predictor of the $K$-means clustering in the cluster $\mathbf{c}_j$ to be :
\begin{equation}
\hat{\mathbf{y}}(\mathbf{\mathbf{c}}_j) = \max\{ F_j(y_1), \cdots, F_j(y_k) \},
\end{equation}
where $F_j(y_i), \cdots, F_j(y_k)$ are the appearance frequencies of each label in the cluster $\mathbf{c}_j$. Then the clustering accuracy can be defined as:
\begin{equation}
\text{Accuracy} = \frac{\sum_{i} 1_{ \{ y_i = \hat{y}_i \} } } {n},
\end{equation}
where $\{\hat{y}_i\}$ are predicted labels. 
Moreover, other evaluation metrics such as precision, recall, and receiver operating characteristic (ROC) can also be defined accordingly.

%===========================================Coil 20=====================================================
\subsubsection{Coil 20}

\begin{figure}[ht]
	\centering
	\includegraphics[width = \textwidth]{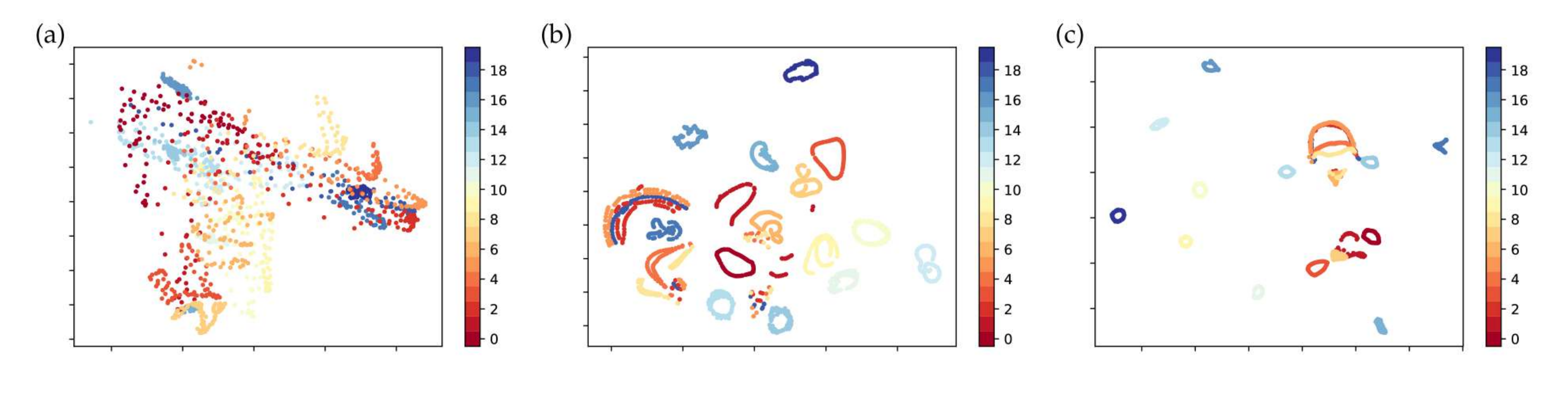}
	\caption{Comparison of different dimensional reduction algorithms on Coil 20 dataset. Total 20 different labels are in the Coil 20 dataset, and we use the ground truth label to color each data points. (a) Feature size is reduced to dimension 2 by PCA. (b) Feature size is reduced to dimension 2 by t-SNE. (c) Feature size is reduced to dimension 2 by UMAP. }
	\label{fig:Coil20}
\end{figure}

\autoref{fig:Coil20} shows the performance of PCA-assisted, UMAP-assisted and t-SNE-assisted clustering of the Coil 20 dataset. For each case, the dataset were reduced to dimension 2 using default parameters, and the plots were colored with the ground truth of the Coil 20 dataset. It can be seen that PCA does not present good  clustering, whereas UMAP and t-SNE show very good clusters. 

\begin{table}[ht]
	\centering
	\caption{Accuracy of $k$-NN of the Coil 20 dataset without applying any reduction algorithms, as well as the accuracy of $k$-NN assisted by PCA, UMAP and t-SNE with different dimensional reduction ratio. The sample size, feature size, and the number of labels of the Coil 20 dataset are 1440, 16384, and 20, respectively. }
	\begin{tabular}{c|cccccc}\hline 
		Dataset & \tabincell{c}{$k$-NN accuracy \\ w/o reduction} & \tabincell{c}{Reduced\\ dimension} & \tabincell{c}{PCA\\accuracy} & \tabincell{c}{UMAP\\accuracy} & \tabincell{c}{t-SNE\\accuracy} \\\hline
		\multirow{10}*{\shortstack{Coil 20 \\(1440,16384,20)}} & \multirow{10}{*}{0.956}
		& 720 (1/2) & 0.955  & 0.668 & 0.850 \\
		&  & 360 (1/4) & 0.957  & 0.861 & 0.889 \\
		&  & 180 (1/8) & 0.973  & 0.867 & 0.881 \\
		&  & 90 (1/16)& 0.977 & 0.860 & 0.885 \\
		&  & 45 (1/32) & 0.980  & 0.861 & 0.875 \\
		&  & 22 (1/64) & 0.985  & 0.868 & 0.743 \\
		&  & 14 (1/100) & 0.730  & 0.851 & 0.878 \\
		&  & 7 (1/200) & 0.985  & 0.870 & 0.845 \\
		&  & 3          & 0.850  & 0.863 & 0.959 \\
		&  & 2          & 0.730  & 0.853 & 0.948 \\ \hline
	\end{tabular}
	\label{tab:Coil20 KNN ACC}
\end{table}

\autoref{tab:Coil20 KNN ACC} shows the accuracy of $k$-NN clustering of the Coil 20 dataset assisted by PCA, t-SNE, and UMAP with different dimensional reduction radio. The Coil 20 dataset   has 1,440 samples, 16,384 features, and 20 different labels. For PCA, the sklearn implementation on python was used with standard parameters. Note that for all methods, dimensions were reduced to 3 and 2 for a comparison. For t-SNE, Multicore-TSNE \cite{Ulyanov2016} was used because it offers up to 8 core processor, which is not available in the sklearn implementation, and it is the fastest performing t-SNE algorithm. For UMAP, we used standard parameters \cite{mcinnes2018umap}. It can be seen that when we reduce the dimension to 3, t-SNE performs best. Moreover, when the dimensional reduction ratio is 1/100, PCA and UMAP also perform well. Notably, the $k$-NN accuracy for the data without applying any dimensional reduction algorithm is 0.956, indicating that UMAP does not provide the best clustering performance on the Coil 20 dataset. However, PCA and t-SNE will preserve the information of the original data with a dimensional reduction ratio larger than 1/100, and t-SNE even performs better for dimensional three on the Coil 20 dataset. 

\begin{table}[ht]
	\centering
	\caption{Accuracy of $K$-means clustering of the Coil 20 dataset without applying any reduction algorithms, as well as the accuracy of $K$-means assisted by PCA, UMAP and t-SNE with different dimensional reduction ratio. The sample size, feature size, and the number of labels of the Coil 20 dataset are 1440, 16384, and 20, respectively. }
	\begin{tabular}{c|cccccc} \hline
		Dataset & \tabincell{c}{$K$-means accuracy \\ w/o reduction} & \tabincell{c}{Reduced\\ dimension} & \tabincell{c}{PCA\\accuracy} & \tabincell{c}{UMAP\\accuracy} & \tabincell{c}{t-SNE\\accuracy} \\\hline
		\multirow{11}*{\shortstack{Coil 20 \\(1440,16384,20)}} & \multirow{11}{*}{0.626} 
		  %  &  16384 (1/1) & 0.626 & NA   & NA     \\ 
		 & 720 (1/2)  & 0.64   & 0.301 & 0.798     \\
		& & 360 (1/4)  & 0.678  & 0.800 & 0.718     \\
		& & 180 (1/8)  & 0.633  & 0.822 & 0.648     \\
		& & 90 (1/16) & 0.642  & 0.799 & 0.681     \\
	    & & 45 (1/32)  & 0.666  & 0.800 & 0.615     \\
		& & 22 (1/64)  & 0.673  & 0.819 & 0.151     \\
		& & 14 (1/100) & 0.631  & 0.817 & 0.154     \\
		& & 7 (1/200)  & 0.591  & 0.819 & 0.360     \\
	    & & 3           & 0.561  & 0.800 & 0.780  \\
		& & 2           & 0.537  & 0.801 & 0.828  \\ \hline
	\end{tabular}
	\label{tab:Coil20 KMean ACC}
\end{table}
\autoref{tab:Coil20 KMean ACC} describes the accuracy of $K$-means clustering of Coil 20 assisted by PCA, UMAP, and t-SNE with different dimensional reduction ratio. For consistency, we use the same set of standard parameters as $k$-NN. For the Coil 20 dataset, the accuracy of $K$-means clustering assisted by UMAP has the best performance. When the reduced dimension is 2048 (ratio 1/8), UMAP will result in a relatively high $K$-means accuracy (0.822). Moreover, although PCA performs best on $k$-NN accuracy, it performs poorly on the $K$-means accuracy, indicating that PCA is not a suitable dimensional reduction algorithm on the Coil 20 dataset. Furthermore, the highest accuracy of $K$-means clustering is 0.828, which is calculated from the t-SNE-assisted algorithm. However, the t-SNE-assisted accuracy under different reduction ratio changes dramatically. When the ratio is 1/64, the t-SNE-assisted accuracy is only 0.151, indicating that t-SNE is sensitive to the hyper-parameters settings. In contrast, the performance of UMAP is highly stable under all dimension-reduction ratios. 

Note that dimension-reduced $k$-means clustering methods outperform the original  $k$-means clustering. Therefore, the proposed dimension-reduced $k$-means clustering methods not only improve the $k$-means clustering efficiency, but also achieve better accuracy.

%=============================================Facebook==================================================
\subsubsection{Facebook Network}
\begin{figure}[ht]
    %facebook
    \centering
	\includegraphics[width = \textwidth]{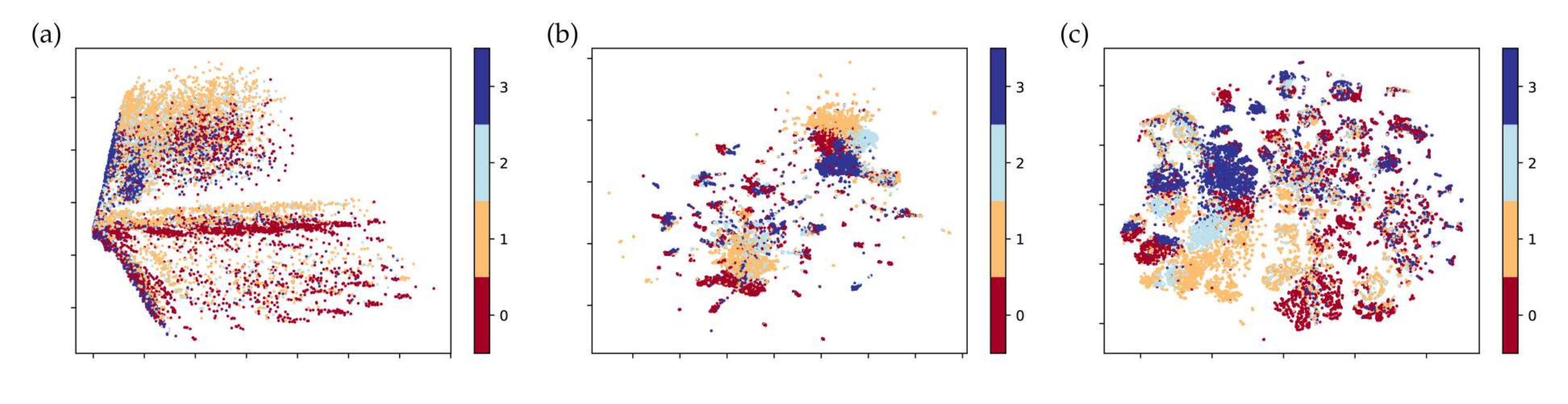}
	\caption{Comparison of different dimensional reduction algorithms on the Facebook Network dataset. Total 4 different labels are in the Facebook Network dataset, and we use the ground truth label to color each data points. (a) Feature size is reduced to dimension 2 by PCA. (b) Feature size is reduced to dimension 2 by t-SNE. (c) Feature size is reduced to dimension 2 by UMAP. }
	\label{fig:Facebook}
\end{figure}
\autoref{fig:Facebook} shows the visualization performance of PCA-assisted, UMAP-assisted, and t-SNE-assisted clustering of the Facebook Network. For each case, the dataset was reduced to dimension 2  using default parameters, and the plots were colored with the ground truth of the Facebook Network. \autoref{fig:Facebook} shows that the PCA-based data is located distributively, while the t-SNE- and UMAP-based data show clusters.

\autoref{tab:Facebook KNN ACC} shows the accuracy of $k$-NN clustering of the Facebook Network assisted by PCA, t-SNE, and UMAP with different dimensional reduction radio. The Facebook Network dataset has 22,470 samples with 4 different labels, and the feature size of the Facebook Network is also 22,470. For each algorithm, we use the same settings as the Coil 20 dataset. Without applying any dimensional reduction method, The Facebook Network has 0.755 $k$-NN accuracy. The reduced feature from PCA has the best $k$-NN performance when the reduction ratio is 1/2. UMAP has a better performance compared to PCA and t-SNE when the reduction ratio is smaller than 1/16. 

\begin{table}[ht!]
	\centering
	\caption{Accuracy of $k$-NN of the Facebook Network without applying any reduction algorithms, as well as the accuracy of $k$-NN assisted by PCA, UMAP and t-SNE with different dimensional reduction ratio. The sample size, feature size, and the number of labels of the Facebook Network are 22470, 22470, and 4, respectively. }
	\begin{tabular}{c|cccccc}\hline 
% 		\multirow{2}{*}{Dataset} & Number & Original  & Reduced & PCA  & UMAP & t-SNE\\
% 		& label & dimension & dimension & accuracy & accuracy & accuracy \\ \hline
        Dataset & \tabincell{c}{$k$-NN accuracy \\ w/o reduction} & \tabincell{c}{Reduced\\ dimension} & \tabincell{c}{PCA\\accuracy} & \tabincell{c}{UMAP\\accuracy} & \tabincell{c}{t-SNE\\accuracy} \\\hline
		\multirow{12}*{\shortstack{Facebook Network \\(22470, 22470, 4)}} & \multirow{13}{*}{0.755}
		  %   & 22470 (1/1) & 0.755  & NA & NA \\
		  & 11235 (1/2) & 0.756 & 0.360 & 0.307 \\
		&  & 5617 (1/4)  & 0.755 & 0.669 & 0.316 \\
		&  & 2808 (1/8)  & 0.754 & 0.754 & 0.355 \\
		&  & 1404 (1/16) & 0.751 & 0.816 & 0.707 \\
		&  & 702 (1/32)  & 0.751 & 0.814 & 0.669 \\
		&  & 351 (1/64)  & 0.746 & 0.815 & 0.690 \\
		&  & 224 (1/100) & 0.733 & 0.814 & 0.676 \\
		&  & 112 (1/200) & 0.721 & 0.819 & 0.633 \\
		&  & 44 (1/500)  & 0.714 & 0.816 & 0.709 \\
		&  & 22 (1/1000) & 0.690 & 0.815 & 0.643 \\
		&  & 3           & 0.552 & 0.801 & 0.741 \\
		&  & 2           & 0.501 & 0.786 & 0.732 \\\hline
	\end{tabular}
	\label{tab:Facebook KNN ACC}
\end{table}

\autoref{tab:Facebook KMean ACC} describes the accuracy of $K$-means clustering of the Facebook Network assisted by PCA, UMAP and t-SNE with different dimensional reduction ratio. PCA, UMAP, and t-SNE all have very poor performance, which may be caused by the smaller number of labels. The highest accuracy 0.427 is observed in the t-SNE-assistant algorithm with dimension 2.

\begin{table}[ht!]
	\centering
	\caption{Accuracy of $K$-means clustering of the Facebook Network without applying any reduction algorithms, as well as the accuracy of $K$-means assisted by PCA, UMAP and t-SNE with different dimensional reduction ratio. The sample size, feature size, and the number of labels of the Facebook Network are 22470, 22470, and 4, respectively.}
	\begin{tabular}{c|cccccc} \hline
% 		\multirow{2}{*}{Dataset}  & Number & Original & Reduced  & PCA  & UMAP  & t-SNE  \\
% 		& labels & dimension & dimension & accuracy & accuracy & accuracy \\ \hline
        Dataset & \tabincell{c}{$K$-means accuracy \\ w/o reduction} & \tabincell{c}{Reduced\\ dimension} & \tabincell{c}{PCA\\accuracy} & \tabincell{c}{UMAP\\accuracy} & \tabincell{c}{t-SNE\\accuracy} \\\hline
		\multirow{12}*{\shortstack{Facebook Network \\(22470, 22470, 4)}} &
		\multirow{12}{*}{0.374}
		  %  & 22470 (1/1) & 0.374 & NA    & NA    \\ 
		& 11235 (1/2) & 0.331 & 0.306 & 0.306    \\
		& & 5617 (1/4)  & 0.331 & 0.307 & 0.299   \\
		& & 2808 (1/8)  & 0.331 & 0.411 & 0.314    \\
		& & 1404 (1/16) & 0.331 & 0.397 & 0.313    \\
		& & 702 (1/32)  & 0.331 & 0.401 & 0.306    \\
		& & 351 (1/64)  & 0.331 & 0.400 & 0.308    \\
		& & 224 (1/100) & 0.331 & 0.400 & 0.327    \\
		& & 112 (1/200) & 0.331 & 0.400 & 0.306    \\
		& & 44 (1/500)  & 0.331 & 0.400 & 0.313    \\
		& & 22 (1/1000) & 0.331 & 0.401 & 0.306    \\
		& & 3           & 0.332 & 0.351 & 0.344 \\
		& & 2           & 0.358 & 0.345 & 0.427 \\ \hline
	\end{tabular}
	\label{tab:Facebook KMean ACC}
\end{table}

Similar to the last case, UMAP-based and t-SNE-based dimension-reduced $k$-means clustering methods outperform the original $k$-means clustering with the full feature dimension. Therefore, it is useful to carry out dimension reduction before $k$-means clustering  for large datasets.

%===============================================MNIST=================================================
\subsubsection{MNIST}

\begin{figure}[ht]
    %MNIST
    \centering
	\includegraphics[width = \textwidth]{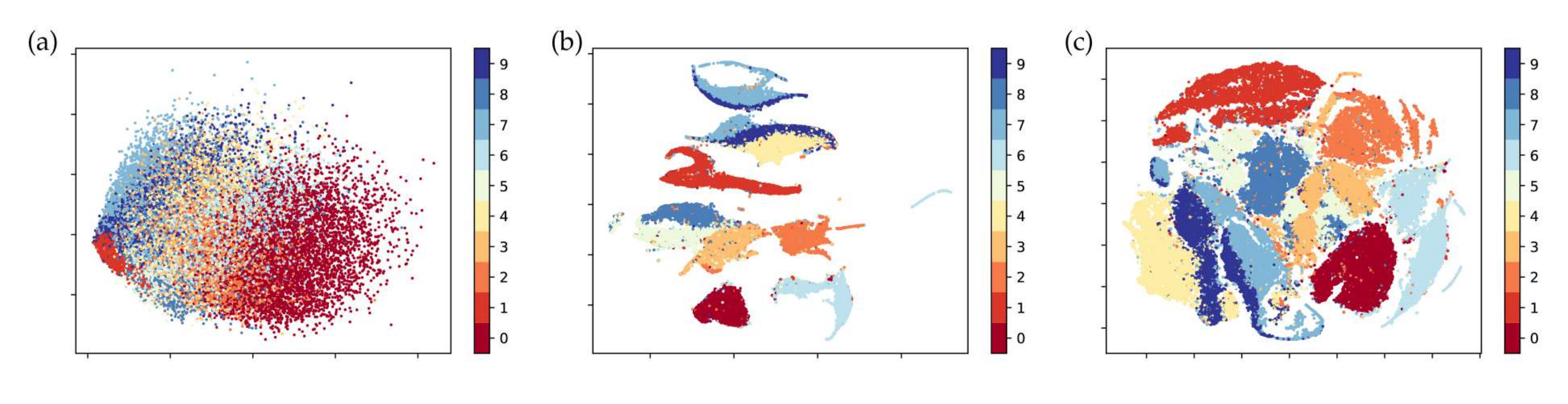}
	\caption{Comparison of different dimensional reduction algorithms on the MNIST dataset. Total 10 different labels are in the MNIST dataset, and we use the ground truth label to color each data points. (a) Feature size is reduced to dimension 2 by PCA. (b) Feature size is reduced to dimension 2 by t-SNE. (c) Feature size is reduced to dimension 2 by UMAP. }
	\label{fig:MNIST}
\end{figure}

\begin{table}[ht!]
	\centering
	\caption{Accuracy of $k$-NN of the MNIST dataset without applying any reduction algorithms, as well as the accuracy of $k$-NN assisted by PCA, UMAP and t-SNE with different dimensional reduction ratio. The sample size, feature size, and the number of labels of the MNIST dataset are 70000, 784, and 10, respectively.}
	\begin{tabular}{c|cccccc}\hline 
% 	    \multirow{2}{*}{Dataset}  & Number & Original & Reduced  & PCA  & UMAP  & t-SNE  \\
% 		& labels & dimension & dimension & accuracy & accuracy & accuracy \\ \hline
        Dataset & \tabincell{c}{$k$-NN accuracy \\ w/o reduction} & \tabincell{c}{Reduced\\ dimension} & \tabincell{c}{PCA\\accuracy} & \tabincell{c}{UMAP\\accuracy} & \tabincell{c}{t-SNE\\accuracy} \\\hline
		\multirow{9}*{\shortstack{MNIST \\(70000, 784, 10)}}   & \multirow{9}{*}{0.948}  
		  %   & 784 (1/1) & 0.948 & NA & NA \\
		& 392 (1/2) & 0.951 & 0.937 & 0.696 \\
		& & 196 (1/4) & 0.956 & 0.938 & 0.846 \\
		& & 98 (1/8)  & 0.960 & 0.937 & 0.893\\
		& & 49 (1/16) & 0.961 & 0.937 & 0.886 \\
		& & 24 (1/32) & 0.953 & 0.937 & 0.842 \\
		& & 12 (1/64) & 0.926 & 0.937 & 0.676 \\
		& & 7 (1/100) & 0.846 & 0.936 & 0.940 \\
		& & 3         & 0.513 & 0.929 & 0.938 \\
		& & 2         & 0.323 & 0.919 & 0.928 \\ \hline
	\end{tabular}
	\label{tab:MNIST KNN ACC}
\end{table}

\autoref{fig:MNIST} shows the performance of PCA-assisted, UMAP-assisted and t-SNE-assisted clustering of the MNIST dataset. The sample size of the MNIST dataset is 70000, which has 784 features with 10 different digit labels. For each case, the dataset was reduced to dimension 2   using default parameters, and the plots were colored with the ground truth of the MNIST dataset. In \autoref{fig:MNIST}, by applying the UMAP algorithm, the clear clusters can be detected for the MNIST dataset. The t-SNE offers a reasonable clustering at  dimension 2 too. However, the PCA does not provide a good clustering.

\begin{table}[ht]
	\centering
	\caption{Accuracy of $K$-means clustering of the MNIST dataset without applying any reduction algorithms, as well as the accuracy of $K$-means assisted by PCA, UMAP and t-SNE with different dimensional reduction ratio. The sample size, feature size, and the number of labels of the MNIST dataset are 70000, 784, and 10, respectively.}
	\begin{tabular}{c|cccccc} \hline
% 		\multirow{2}{*}{Dataset}  & Number & Original & Reduced  & PCA  & UMAP  & t-SNE  \\
% 		& labels & dimension & dimension & accuracy & accuracy & accuracy \\ \hline
        Dataset & \tabincell{c}{$K$-means accuracy \\ w/o reduction} & \tabincell{c}{Reduced\\ dimension} & \tabincell{c}{PCA\\accuracy} & \tabincell{c}{UMAP\\accuracy} & \tabincell{c}{t-SNE\\accuracy} \\\hline
		\multirow{9}*{\shortstack{MNIST \\(70000, 784, 10)}} & \multirow{9}{*}{0.494}
		  %  & 784 (1/1) & 0.494 & NA    & NA   \\
	    & 392 (1/2) & 0.487 & 0.665 & 0.122   \\
		& & 196 (1/4) & 0.492 & 0.667 & 0.113   \\
		& & 98 (1/8)  & 0.498 & 0.673 & 0.113   \\
		& & 49 (1/16) & 0.496 & 0.718 & 0.113   \\
		& & 24 (1/32) & 0.501 & 0.697 & 0.114   \\
		& & 12 (1/64) & 0.489 & 0.682 & 0.138   \\
		& & 7 (1/100) & 0.464 & 0.677 & 0.740   \\
		& & 3         & 0.365 & 0.727 & 0.537 \\
		& & 2         & 0.300 & 0.712 & 0.593 \\ \hline
	\end{tabular}
	\label{tab:MNIST KMean ACC}
\end{table}

\autoref{tab:MNIST KNN ACC} shows the accuracy of $k$-NN clustering of the MNIST dataset assisted by PCA, t-SNE, and UMAP with different dimensional reduction radios. For each algorithm, we use the same settings as the Coil 20 dataset. Without applying any dimensional reduction algorithms, the accuracy of $k$-NN is 0.948. By applying PCA/UMAP with the reduction ratio greater than 1/64, the accuracy of PCA/UMAP-assisted $k$-NN is at the same level without using any dimensional reduction algorithm. However, in contract with UMAP and t-SNE, when the reduced dimension is 2 or 3, PCA performs poorly. This indicates that the PCA may  not be suitable for dimension-reduction for  datasets with a large sample size.

\autoref{tab:MNIST KMean ACC} describes the accuracy of $K$-means clustering of the MNIST dataset assisted by PCA, UMAP, and t-SNE with different dimensional reduction ratios.  By applying PCA, the accuracy of $K$-means is around 0.45. 
The t-SNE method performance is quite unstable, from very poor (0113) to the best (0.740), and to a relatively low value of 0.593. In contrast, we can see a stable and improved accuracy from using UMAP at various reduction ratios, indicating that the reduced feature generated by UMAP can better represent the clustering properties of the MNIST dataset compared to the PCA and t-SNE. 

As observed early, the present UMAP and t-SNE-assisted $k$-means clustering methods also significantly out-perform the original $k$-means clustering for this dataset.

%=====================================Jaccard distanced-basedMNIST==================================================
\subsubsection{Jaccard distanced-based MNIST}

\begin{figure}[ht]
	\centering
	%Jaccard distanced-based MNIST
	\includegraphics[width = \textwidth]{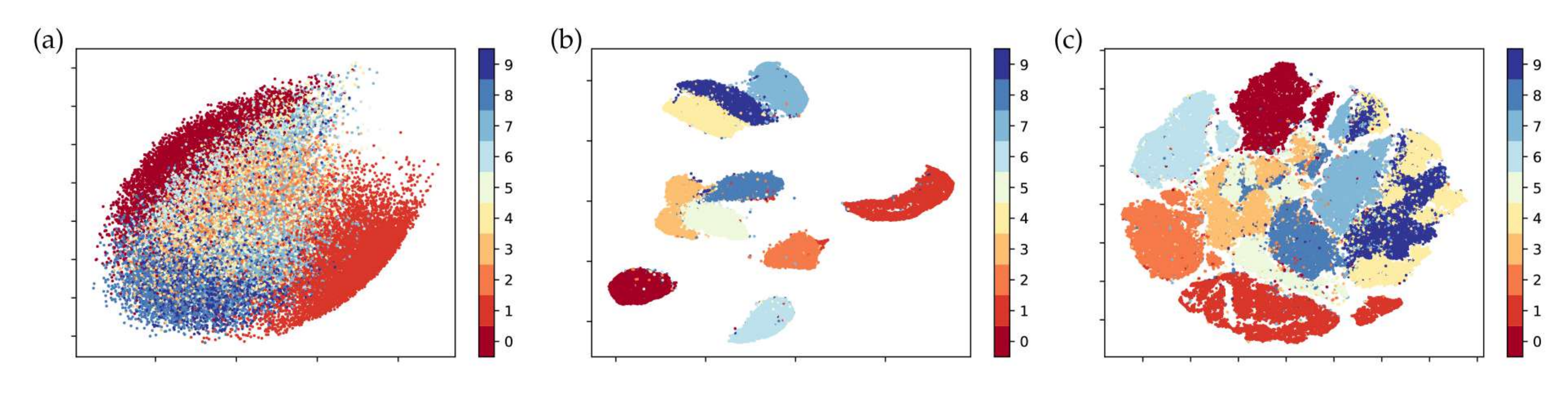}
	\caption{Comparison of different dimensional reduction algorithms on the Jaccard distanced-based MNIST dataset. Total 10 different labels are in the Jaccard distanced-based MNIST dataset, and we use the ground truth label to color each data points. (a) Feature size is reduced to dimension 2 by PCA. (b) Feature size is reduced to dimension 2 by t-SNE. (c) Feature size is reduced to dimension 2 by UMAP. }
	\label{fig:Jaccard distanced-based MNIST}
\end{figure}

Out last validation dataset is Jaccard distanced-based MNIST. This dataset can be treated as a test on the Jaccard distance-based data representation. 
\autoref{fig:Jaccard distanced-based MNIST} shows the performance of PCA-assisted, UMAP-assisted, and t-SNE-assisted clustering of the Jaccard distanced-based MNIST dataset. The dataset was reduced to dimension 2  using default parameters for visualization, and the plots were colored with the ground truth of the Jaccard distanced-based MNIST dataset. From \autoref{fig:Jaccard distanced-based MNIST}, we can see that UMAP provides the clearest clusters compared to PCA and t-SNE when the dimension is reduced to 2. The performance of t-SNE is reasonable while PCA does not give a good clustering.   

\begin{table}[ht!]
	\centering
	\caption{Accuracy of $k$-NN of the Jaccard distanced-based MNIST dataset without applying any reduction algorithms, as well as the accuracy of $k$-NN assisted by PCA, UMAP and t-SNE with different dimensional reduction ratio. The sample size, feature size, and the number of labels of the Jaccard distanced-based MNIST dataset are 70000, 70000, and 10, respectively.}
	\begin{tabular}{c|cccccc}\hline 
% 	    \multirow{2}{*}{Dataset}  & Number & Original & Reduced  & PCA  & UMAP  & t-SNE  \\
% 		& labels & dimension & dimension & accuracy & accuracy & accuracy \\ \hline
        Dataset & \tabincell{c}{$k$-NN accuracy \\ w/o reduction} & \tabincell{c}{Reduced\\ dimension} & \tabincell{c}{PCA\\accuracy} & \tabincell{c}{UMAP\\accuracy} & \tabincell{c}{t-SNE\\accuracy} \\\hline
		\multirow{13}*{\shortstack{Jaccard distanced-based MNIST \\(70000, 70000, 10)}} & \multirow{13}{*}{0.958} 
		      %& 70000 (1/1) & 0.958 & NA & NA \\
		& 7000 (1/10) & 0.958 & 0.958 & NA  \\
		&  & 3500 (1/20) & 0.958 & 0.966 & NA \\
		&  & 1750 (1/40) & 0.958 & 0.967 & NA \\
		&  & 875 (1/80)  & 0.958 & 0.967 & NA \\ 
		&  & 437 (1/160) & 0.958 & 0.968 & 0.718 \\
		&  & 218 (1/320) & 0.958 & 0.968 & 0.701 \\
		&  & 109 (1/640) & 0.958 & 0.968 & 0.873 \\
		&  & 70 (1/1000) & 0.958 & 0.968 & 0.915 \\
		&  & 35 (1/2000) & 0.956 & 0.968 & 0.872 \\
		&  & 17 (1/5000) & 0.938 & 0.968 & 0.916 \\
		&  & 7 (1/10000) & 0.867 & 0.967 & 0.942 \\ 
		&  & 3           & 0.487 & 0.965 & 0.939 \\ 
	    &  & 2           & 0.313 & 0.960    & 0.924 \\\hline
	\end{tabular}
	\label{tab:Jaccard distanced-based MNIST KNN ACC}
\end{table}

\begin{table}[ht!]
	\centering
	\caption{Accuracy of $K$-means clustering of the Jaccard distanced-based MNIST dataset without applying any reduction algorithms, as well as the accuracy of $K$-means assisted by PCA, UMAP and t-SNE with different dimensional reduction ratio. The sample size, feature size, and the number of labels of the Jaccard distanced-based MNIST dataset are 70000, 70000, and 10, respectively.}
	\begin{tabular}{c|cccccc} \hline
% 		\multirow{2}{*}{Dataset}  $K$-means accuracy & Reduced  & PCA  & UMAP  & t-SNE  \\
% 		& labels & dimension & dimension & accuracy & accuracy & accuracy \\ \hline
        Dataset & \tabincell{c}{$K$-means accuracy \\ w/o reduction} & \tabincell{c}{Reduced\\ dimension} & \tabincell{c}{PCA\\accuracy} & \tabincell{c}{UMAP\\accuracy} & \tabincell{c}{t-SNE\\accuracy} \\\hline
		\multirow{13}*{\shortstack{Jaccard distanced-based MNIST \\(70000, 70000, 10)}} & \multirow{13}{*}{0.555}
		  %  & 70000 (1/1) & 0.555 & NA    & NA \\ 
		& 7000 (1/10) & 0.436 & 0.329 & NA   \\
		& & 3500 (1/20) & 0.436 & 0.693 & NA   \\
		& & 1750 (1/40) & 0.436 & 0.792 & NA   \\
		& & 875 (1/80)  & 0.435 & 0.793 & NA   \\
		& & 437 (1/160) & 0.435 & 0.793 & 0.114   \\
		& & 218 (1/320) & 0.435 & 0.793 & 0.156   \\
		& & 109 (1/640) & 0.435 & 0.794 & 0.114   \\
		& & 70 (1/1000) & 0.436 & 0.793 & 0.113  \\
		& & 35 (1/2000) & 0.435 & 0.794 & 0.116   \\
		& & 17 (1/5000) & 0.436 & 0.793 & 0.113   \\
		& & 7 (1/10000) & 0.431 & 0.793 & 0.737   \\
		& & 3           & 0.364 & 0.798 & 0.635 \\
		& & 2           & 0.261 & 0.791 & 0.635 \\\hline
	\end{tabular}
	\label{tab:Jaccard distanced-based MNIST KMean ACC}
\end{table}

\autoref{tab:Jaccard distanced-based MNIST KNN ACC} shows the accuracy of $k$-NN clustering of Jaccard distanced-based MNIST assisted by PCA, t-SNE, and UMAP with different dimensional reduction radios. For each algorithm, we use the same settings as the Coil 20 dataset. Notably, the $k$-NN accuracy for the data without applying any dimensional reduction algorithm is 0.958, which is at the same level as the PCA algorithm with a reduction ratio greater than 1/5000. Moreover, we can find that UMAP performs well compared to PCA and t-SNE, indicating that after applying UMAP, the reduced feature still preserves most of the valued information of the Jaccard distanced-based MNIST dataset. The stability and persistence of 
UMAP at various reduction ratios are the most important features. 

\autoref{tab:Jaccard distanced-based MNIST KMean ACC} describes the accuracy of $K$-means clustering of the Jaccard distanced-based MNIST dataset assisted by PCA, UMAP, and t-SNE with different dimensional reduction ratio. For consistency, we will use the same standard parameters as $k$-NN. Similar to the MNIST dataset, the accuracy of $K$-means clustering assisted by UMAP still has the best performance. When the reduced dimension is 3, UMAP will result in the highest $K$-means accuracy 0.798. Noticeably, although PCA performs well on $k$-NN accuracy, it has the lowest $K$-mean accuracy, indicating that PCA is not a suitable dimensional reduction algorithm, especially for those datasets with a large number of samples. To be noted, the t-SNE accuracy at four reduced dimensions are not available due to the extremely long running time.

In a nutshell, PCA, UMAP, and t-SNE can all perform well for $k$-NN. However, for the Coil 20 dataset, UMAP performs slightly poorly, whereas the t-SNE performs well, which may be caused by a lack of data size. In order to train UMAP, it needs a suitable data size. The Coil 20 dataset has 20 labels, each with only 72 samples. This may not be enough to train UMAP properly. However, even in this case, UMAP performance is still very stable at various reduction ratios and is the best method in terms of reliability, which become the major advantages of UMAP. Another strength of UMAP comes from its dimension-reduction for $K$-means clustering. In most cases, UMAP can improve $K$-means clustering accuracy, especially for the Jaccard distanced-based MNIST dataset. Furthermore, UMAP can generate a very clear and elegant visualization of clusters with low dimensional reduction value such as 2. Additionally, UMAP performed better than PCA and t-SNE for a larger dataset (MNIST and Jaccard distanced-based MNIST). Especially for the Jaccard distanced-based MNIST data, where Jaccard distance was used as the metric, UMAP performed best, which indicates the merit of using UMAP for Jaccard distanced-based datasets, such as COVID-19 SNP datasets. Furthermore, the accuracies for $k$-NN classification and $K$-means clustering are both improved on the Jaccard distance-based MNIST dataset compared to the original MNIST dataset, which provides convincing evidence that the Jaccard distance representation will   help improve the performance of the clustering on the SARS-CoV-2 mutation dataset in the following sections.

\subsection{Efficiency comparison}
 
\begin{figure}[ht]
	\centering
	\begin{subfigure}{.49\textwidth}
		\centering
		\includegraphics[width = \textwidth]{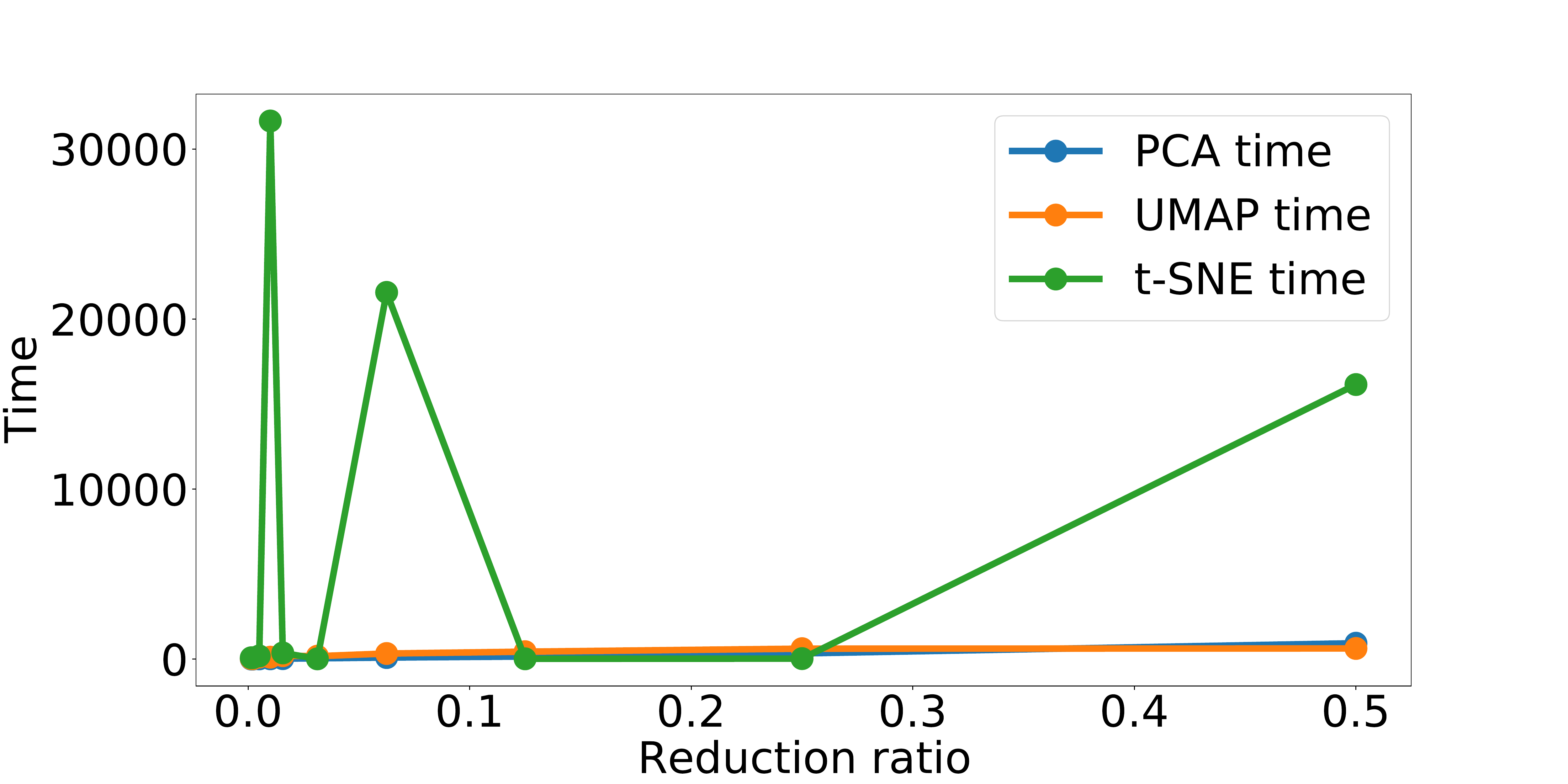}
		\subcaption{Coil 20 time}
	\end{subfigure}
	\begin{subfigure}{.49\textwidth}
		\centering
		\includegraphics[width = \textwidth]{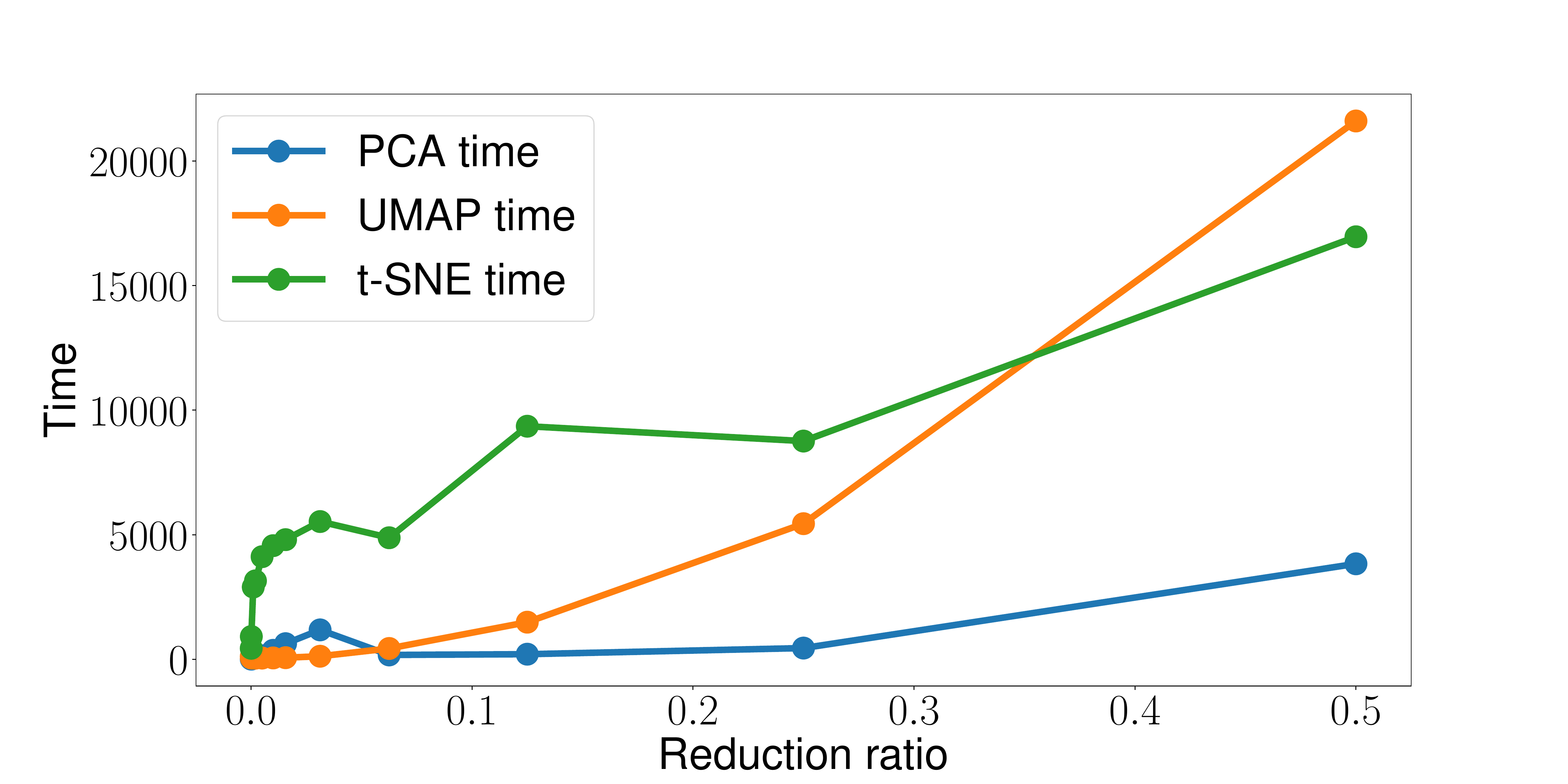}
		\subcaption{Facebook Network time}
	\end{subfigure}
	
	\begin{subfigure}{.49\textwidth}
		\centering
		\includegraphics[width = \textwidth]{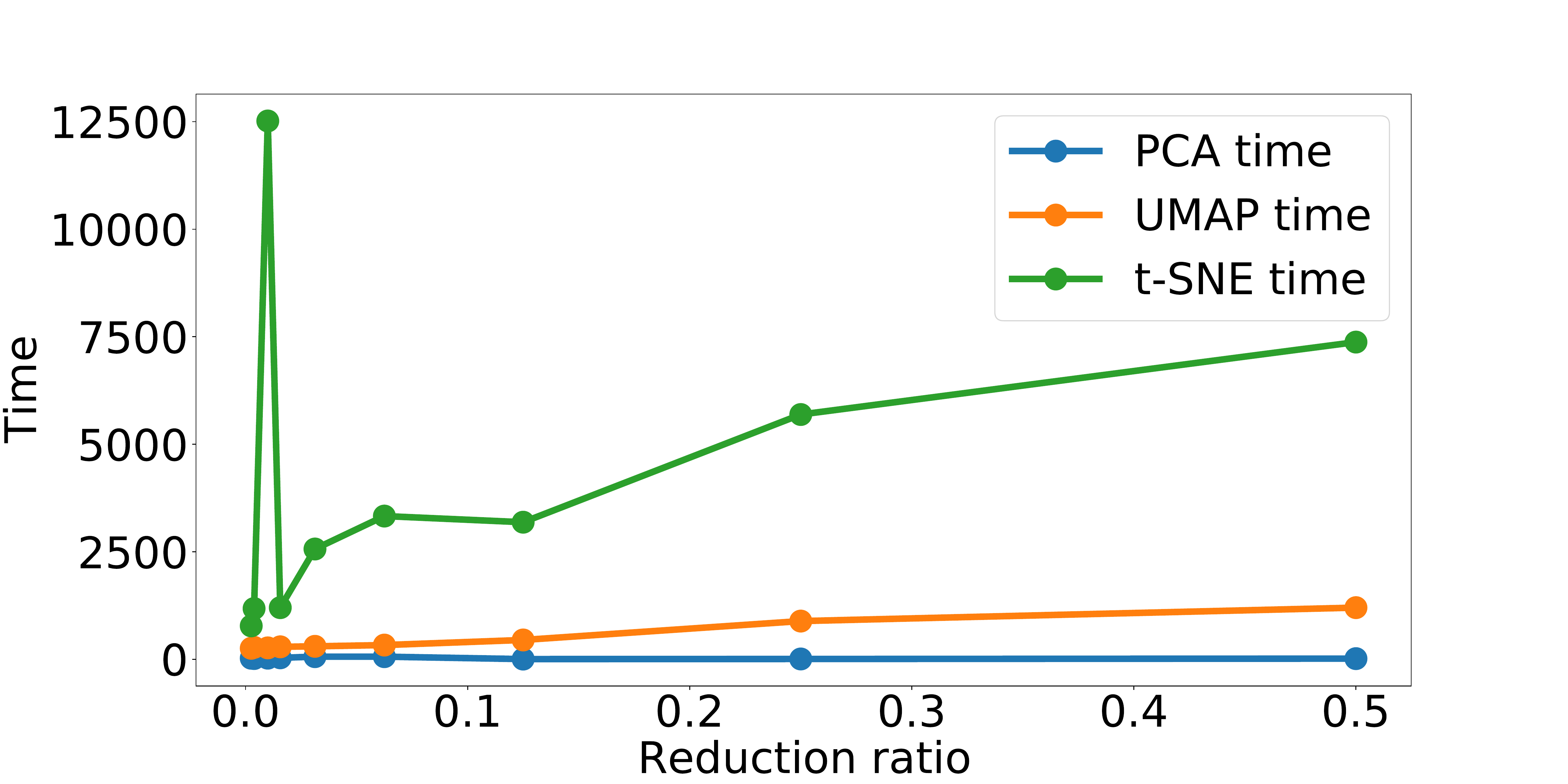}
		\subcaption{MNIST time}
	\end{subfigure}
	\begin{subfigure}{.49\textwidth}
		\centering
		\includegraphics[width = \textwidth]{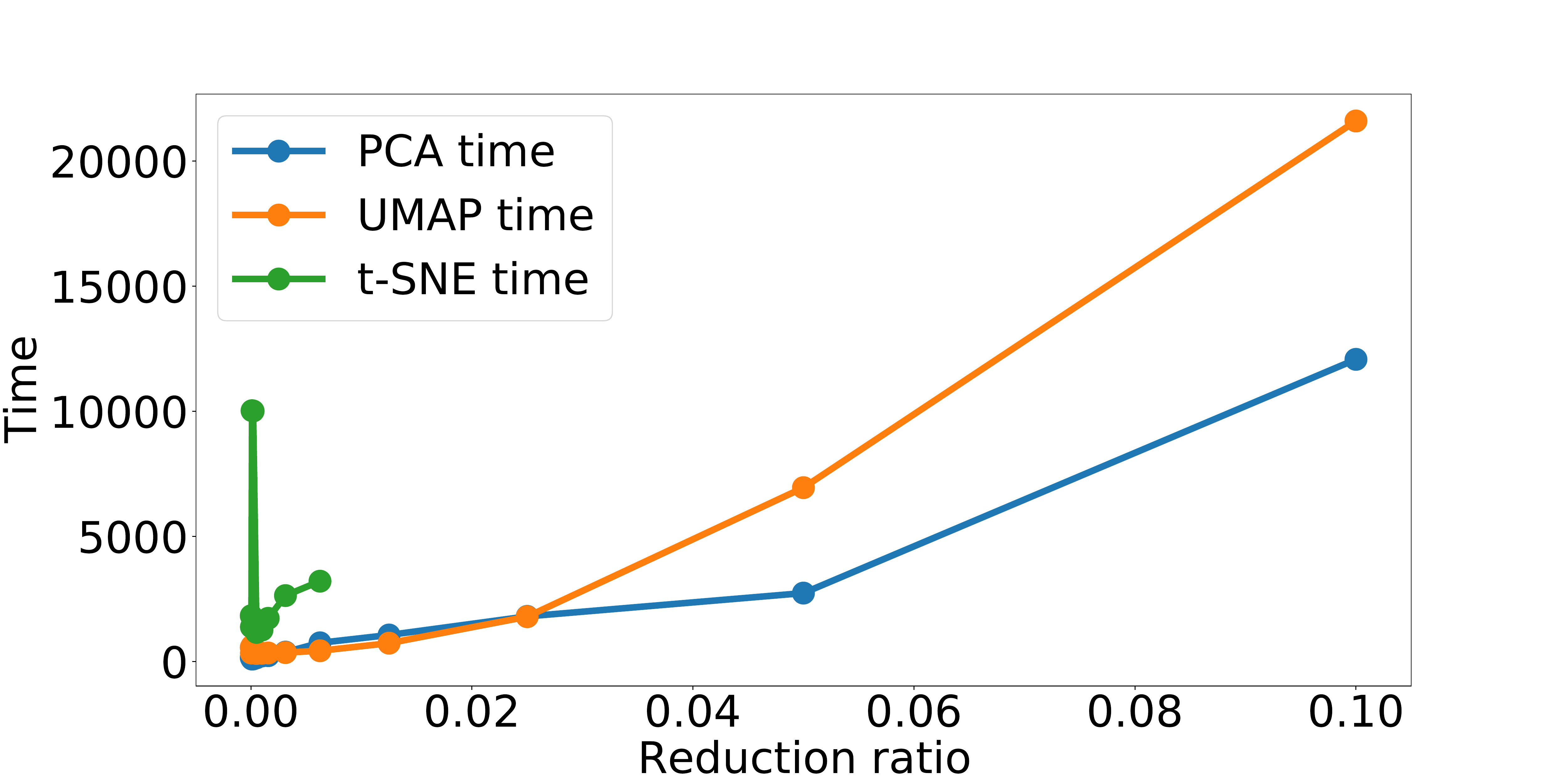}
		\subcaption{Jaccard distanced-based MNIST time}
	\end{subfigure}
	\caption{Computational time of each reduction ratio. The green, orange and blue lines represent the computational time of t-SNE, UMAP, and PCA, respectively. Not surprisingly, PCA performs the best in the majority of cases, except for the Coil 20 dataset. UMAP and t-SNE perform worse than PCA, but UMAP scales better when there are more samples, as evident from MNIST and Jaccard distanced-based MNIST datasets. Note that for Jaccard distanced-based MNIST, the higher dimension was not computed because the computational time was too long.}
	\label{fig:time}
\end{figure}

It is important to understand the computational time behaviors of various methods. To this end, we compare computational time for three dimension-reduction techniques. \autoref{fig:time} depicts the computational time of three methods for the four datasets under various reduction ratios. The green, orange, and blue lines represent the computational time of t-SNE, UMAP, and PCA, respectively. Some points in green line of \autoref{fig:time} (d) are not available, which due to the extremely long running time. PCA performed best in most cases, except for the Coil 20 dataset, where UMAP had comparable computational time. This behavior is expected because PCA is a linear transformation, and its time should scale linearly with the number of components in the lower dimensional space. UMAP and t-SNE were slower than PCA, but it is evident from MNIST and Jaccard distanced-based MNIST datasets that UMAP scales better with the increase in the number of samples. 
Note that for Jaccard distanced-based MNIST, a higher dimension was not computed because the computational time was too long. For Facebook Network, UMAP is outperforming t-SNE; however, for higher dimensions, t-SNE computed faster. Nonetheless, from our baseline test \autoref{tab:Facebook KNN ACC}, t-SNE does not perform well, indicating instability. Faster computation time may indicate too fast of a convergence, which leads to poor embedding.

% \clearpage

\section{SARS-CoV-2 mutation clustering}
% Data collected up to October 30, 2020 were analyzed. 

%Based on the above analysis, we choose UMAP to perform the dimension  reduction of SARS-CoV-2 genome sequence data. 

\subsection{World SARS-CoV-2 mutation clustering}
 We gather data submitted to GISAID up to October 30, 2020, and the total number of samples is 89627. We first get the SNP information by applying the multiple sequence alignment, which leads to 23763 unique SNPs. Next, we calculate the pairwise Jaccard distance of our dataset in order to generate the Jaccard distance-based features. Here, the number of rows is the number of samples (89627), and the number of columns is the feature size (89627). As we mentioned in Section \ref{sec:Jaccard}, the Jaccard distance-based feature is a square matrix. However, due to the large size of samples and features, applying $K$-means clustering directly on the feature of the size of 89627$\times$89627 is a very time-consuming process. Considering that UMAP outperforms the other two dimensional reduction algorithms (PCA and t-SNE) on the Jaccard distance-based MNIST dataset, we  employ UMAP to reduce our original feature with the size of 89627$\times$89627 to 89627$\times$200. To be noted, UMAP is a reliable and stable algorithm, which performs consistently in clustering at various reduction ratios. Therefore, there is no need to use the same reduction dimension of 200 and one can also choose a different reduction dimension value to generate similar results. 

With the reduced dimension feature that has the size of 89627$\times$200, we  split our SARS-CoV-2 dataset into different clusters by applying the $K$-means clustering methods. After comparing the WCSS under a different number of clusters, we find that there are 6 clusters forming within the SARS-CoV-2 population based on the elbow method. \autoref{tab: 1030 mut} shows the top 25 single mutations of each cluster. In order to understand the relationship, we also analyzed the commutation occurring in each cluster (\autoref{tab: 1030 co}). From \autoref{tab: 1030 mut} and \autoref{tab: 1030 co} we see the following:

\begin{table}[ht]
	\centering
	\scriptsize
	\setlength\tabcolsep{1pt}
	\caption{The frequency and occurrence percentage of SARS-CoV-2 co-mutations from each clusters in the world.}
	\begin{tabular}{clcc}\hline 
		Cluster & Co-mutations & Frequency & Occurrence percentage \\\hline
		Cluster 1 & [241, 1163, 3037, 7540, 14408, 16647, 18555, 22992, 23401, 23403, 28881, 28882, 28883] & 776 & 0.463 \\
		Cluster 2 & [241, 3037, 14408, 23403] & 8640 & 0.925\\
		Cluster 3 & [241, 1059, 3037, 14408, 23403, 25563] & 8878 & 0.662 \\
		Cluster 4 & [241, 3037, 14408, 23403, 28881, 28882, 28883] & 14913 & 0.829 \\
		Cluster 5 & [241, 3037, 14408, 23403] & 17412 & 0.969 \\
		Cluster 6 & [241, 1163, 3037, 7540, 14408, 16647, 18555, 22992, 23401, 23403, 28881, 28882, 28883] & 1352 & 0.771 \\ \hline
	\end{tabular}
	\label{tab: 1030 co}
\end{table}

\begin{itemize}
	\item Though Clusters 1 and 6 seem similar from the top 25 single mutations, the co-mutations tells a different story. The same co-mutations have a higher frequency in Cluster 6, indicating that the co-mutation has higher number of descendants. 
	\item Clusters 2 and 5 have high frequency of [241, 3037, 14408, 23403] mutations, but Cluster 5 has a clear co-mutation descendent with high frequency.
	\item Cluster 3 has a unique combination of mutation that is only popular in Cluster 3.
\end{itemize}

\autoref{tab: 1030Country} shows the cluster distributions of samples from 25 countries. Here, we use the ISO 3166-1 alpha-2 codes as the country code. The listed countries are the United Kingdom (UK), the United States (US), Australia (AU), India (IN), Switzerland (CH), Netherlands (NL), Canada (CA), France (FR), Belgium (BE), Singapore (SG), Spain (ES), Russia (RU), Portugal (PT), Denmark (DK), Sweden (SE), Austria (AT), Japan (JP), South Africa (ZA), Iceland (IS), Brazil (BR), Saudi Arabia (SA), Norway (NO), China (CN), Italy (IT), and Korea (KR).  From \autoref{tab: 1030Country}, we can see the following:
\begin{itemize}
	\item SNP profiles from UK are dominated in Clusters 5 and 4.
	\item Clusters 1 and 6's SNP profiles are predominantly found in AU.
	\item SNP profiles from US are found mostly in Clusters 3 and 5.
	\item Most country's SNP profiles are found in Clusters 2 - 5, with some having slightly higher numbers, but not as significant as the UK, US and AU.
\end{itemize}

\begin{figure}[ht]
	\centering
	\includegraphics[width = \textwidth]{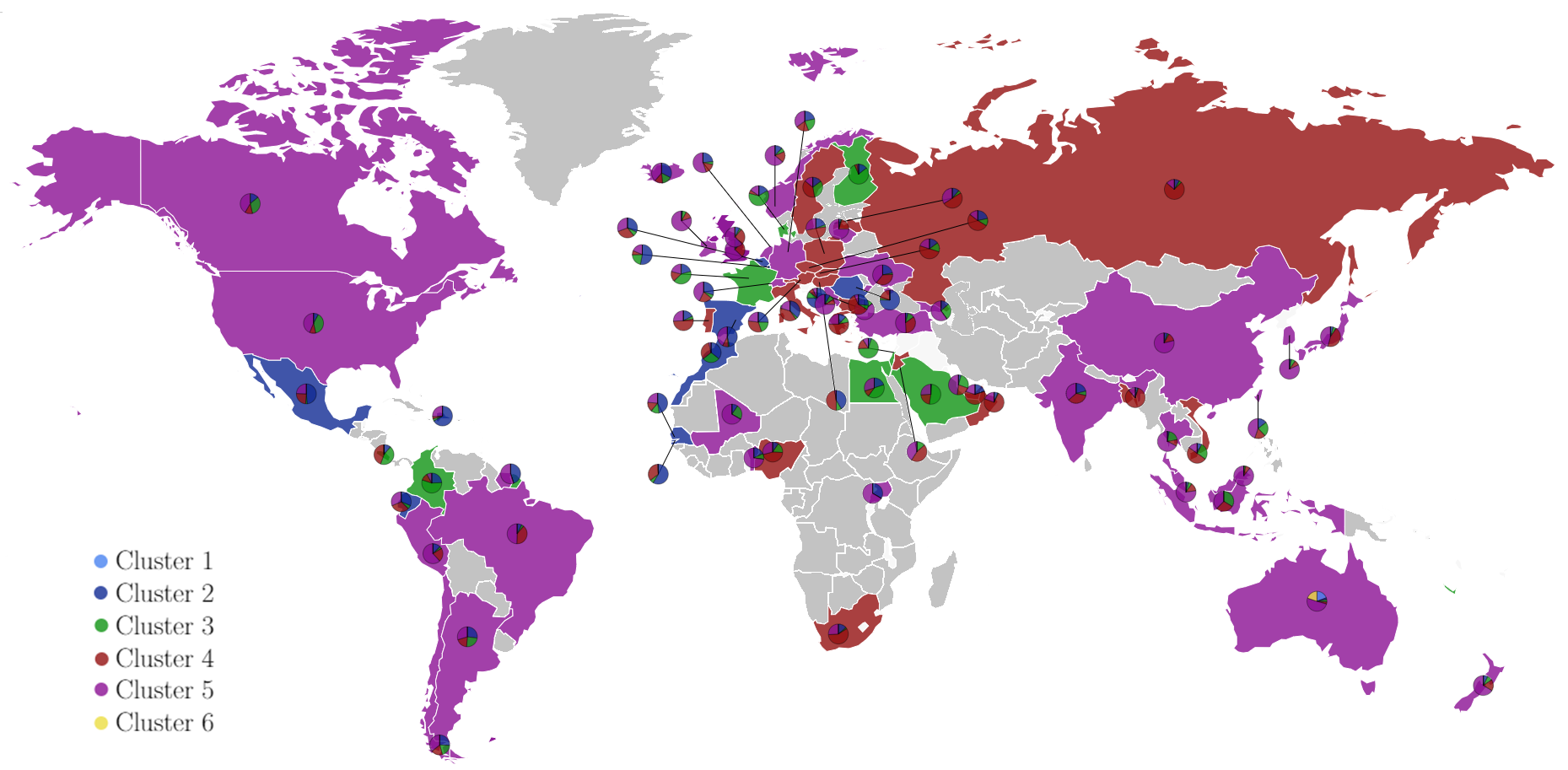}
	\caption{Cluster distribution of the global SARS-CoV-2 mutation dataset. Using Highchart, the world map was colored, according to the dominant cluster. Clusters 1, 2, 3, 4, 5, and 6 were colored with light blue, blue, green, red, purple and yellow, respectively. For example, United States have SNP profiles from all clusters, but Cluster 5 (purple) is the dominant type in the US. Only countries with more than 25 sequenced data available on GISAID were considered. Countries with fewer than 25 samples are labeled grayed.}
\end{figure}

\begin{figure}[ht]
    \centering
    \includegraphics[width = 0.9\textwidth]{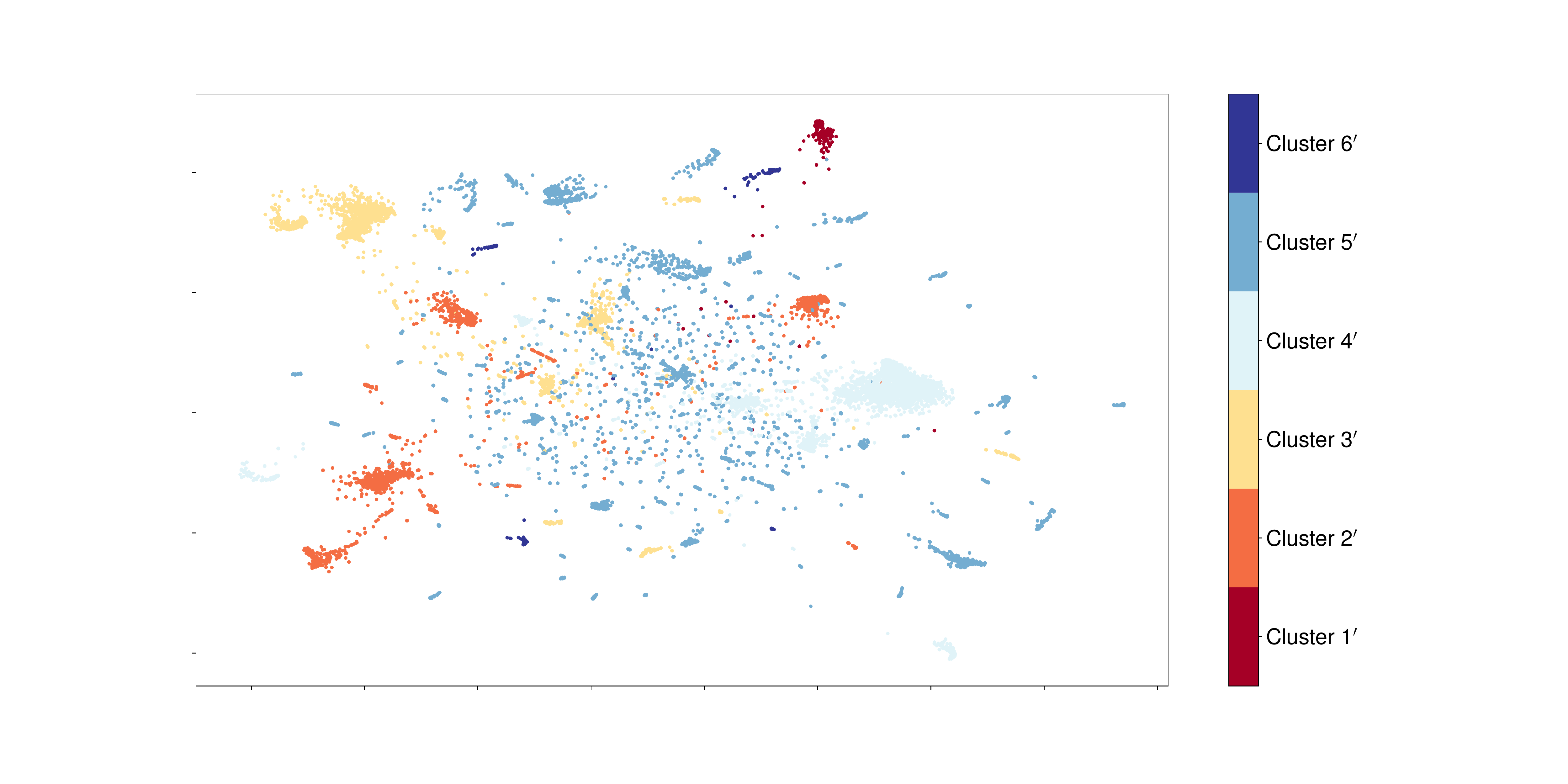}
    \caption{ 2D UMAP visualization of the world  SARS-CoV-2 mutation dataset with 6 distinct clusters. Red, orange, yellow, light blue, blue, and dark blue represent for Clusters 1', 2', 3', 4', 5', and 6', respectively.} 
    \label{fig:1030_World_UMAP}
\end{figure}

  Notably, in \autoref{tab: 1030 co}, Cluster 2 and Cluster 5 have the same co-mutations with a relatively large frequency, while Cluster 1 and Cluster 6 share the same co-mutations with a relatively low frequency, which indicate that Cluster 2 and Cluster 5 share the same ``root" with a large size, while Cluster 1 and Cluster 6 share the same ``root" with a smaller size in the 200-dimensional (200D) space. However, we cannot visualize the distribution of our reduced dataset in the 200D space. Therefore, benefit from the stable and reliable performance of UMAP at various reduction ratios, we reduce the dimension of our original dataset to 2, which enables us to observe the distribution of the dataset in the two-dimensional (2D) space.  \autoref{fig:1030_World_UMAP} visualizes the distribution of our dataset with 6 distinct clusters with 2D UMAP. It can be seen that 2 clusters (i.e., Cluster 2' and Cluster 3') share a small ``root" located in the middle of the figure, and Cluster 4' and Cluster 5' share another large ``root" that also located in the middle of the figure.

\subsection{United States SARS-CoV-2 mutation clustering}
\begin{figure}[ht]
	\centering 
	\includegraphics[width = \textwidth]{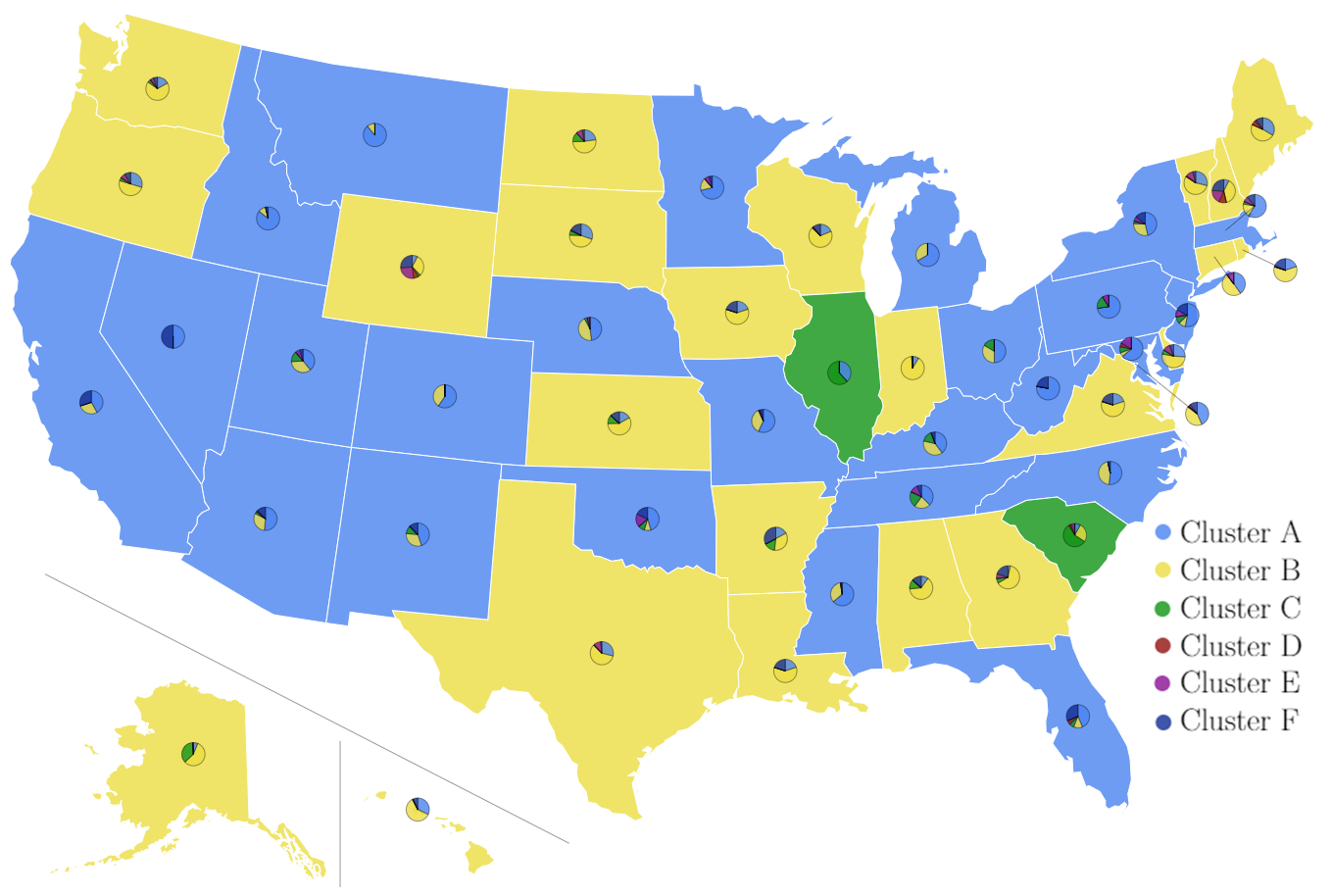}
	\caption{Cluster distribution of United States SARS-CoV-2 mutation dataset. Using Highchart, the US map was colored, according to the dominant cluster. Clusters A, B, C, D, E, and F were colored with light blue, blue, green, red, purple, and yellow, respectively. For example, United States have SNP profiles from all clusters, but Cluster E (purple) is the dominant type in the US. Only those countries that have more than 25 sequenced data available on GISAID were considered in the plot. }
	\label{fig: USMap}
\end{figure}

In addition to analyzing the clustering in the world, SNP profiles of SARS-CoV-2 from the United States (US) were considered. In this section, the US dataset has 10279 unique single mutations and 22390 samples. Therefore, the dimension of the Jaccard distance-based dataset is 22390$\times$22390. After applying the UMAP, we reduce the dimension of the original dataset to be 22390$\times$200. Following the similar $K$-means clustering processes as we did for the world dataset,  we find that there are 6 predominant clusters forming in the United States. \autoref{fig: USMap} show the US map with the cluster statistic. Here, Highchart was used to generate the plot with the pie chart. Each states were colored based on the dominant cluster.

\autoref{tab: 1030US mut} shows the top 25 mutations from each clusters in the United States. The states with more than 50 samples are listed.  \autoref{tab: 1030US co} shows the common occurring co-mutations, and we can observe the following:

% Note that the cluster from (\autoref{tab: 1030 mut}, \autoref{tab: 1030 co}) are different from clusters A, B, C, D, E, and F. 

\begin{itemize}
	\item Cluster F have high frequency of co-mutations [241, 3037, 14408, 23403, 28881, 28882, 28883], which is a descendent of common co-mutations of Cluster 4 [241, 3037, 14408, 23403, 28881, 28882, 28883]  from \autoref{tab: 1030US mut}.
	\item Clusters A, B, C, and D have frequent co-mutations [241, 1059, 3037, 14408, 23403, 25563], which are also   frequent co-mutations of Cluster 3.
% 	\item {\color{red} Cluster A, B, and C share the same high frequency co-mutations [241, 1059, 3037, 14408, 23403, 25563]. However, the occurrence percentages are different. {(I need one figure to explain this. Yuta is working on it.)}}
\end{itemize}

\begin{table}[ht]
	\centering
	\caption{The frequency and occurrence percentage of SARS-CoV-2 co-mutations from each clusters in US clusters.}
	\begin{tabular}{c l c c }\hline 
		Cluster & Co-mutations & Frequency & Occurrence percentage \\\hline
		Cluster A & [241, 1059, 3037, 14408, 23403, 25563] & 3116 & 0.465 \\
		Cluster B & [241, 1059, 3037, 14408, 23403, 25563] & 5763 & 0.605\\
		Cluster C & [241, 1059, 3037, 14408, 23403, 25563] & 8878 & 0.662 \\
		Cluster D & [241, 1059, 3037, 14408, 23403, 25563, 27964] & 1225 & 0.864 \\
		Cluster E & [8782, 17747, 17858, 18060, 28144] & 1109 & 0.743\\
		Cluster F & [241, 3037, 14408, 23403, 28881, 28882, 28883] & 2575 & 0.932 \\ \hline
	\end{tabular}
	\label{tab: 1030US co}
\end{table}

% \autoref{tab: 1030US mut} show the cluster statistic for each states. States with more than 50 samples submitted to GISAID are listed. \autoref{fig: USMap} show the US map with the cluster statistic. Highchart was used to generate the plot with the pie chart. Each states were colored based on the dominant cluster.

\begin{figure}[ht]
    \centering
    \includegraphics[width = 0.9\textwidth]{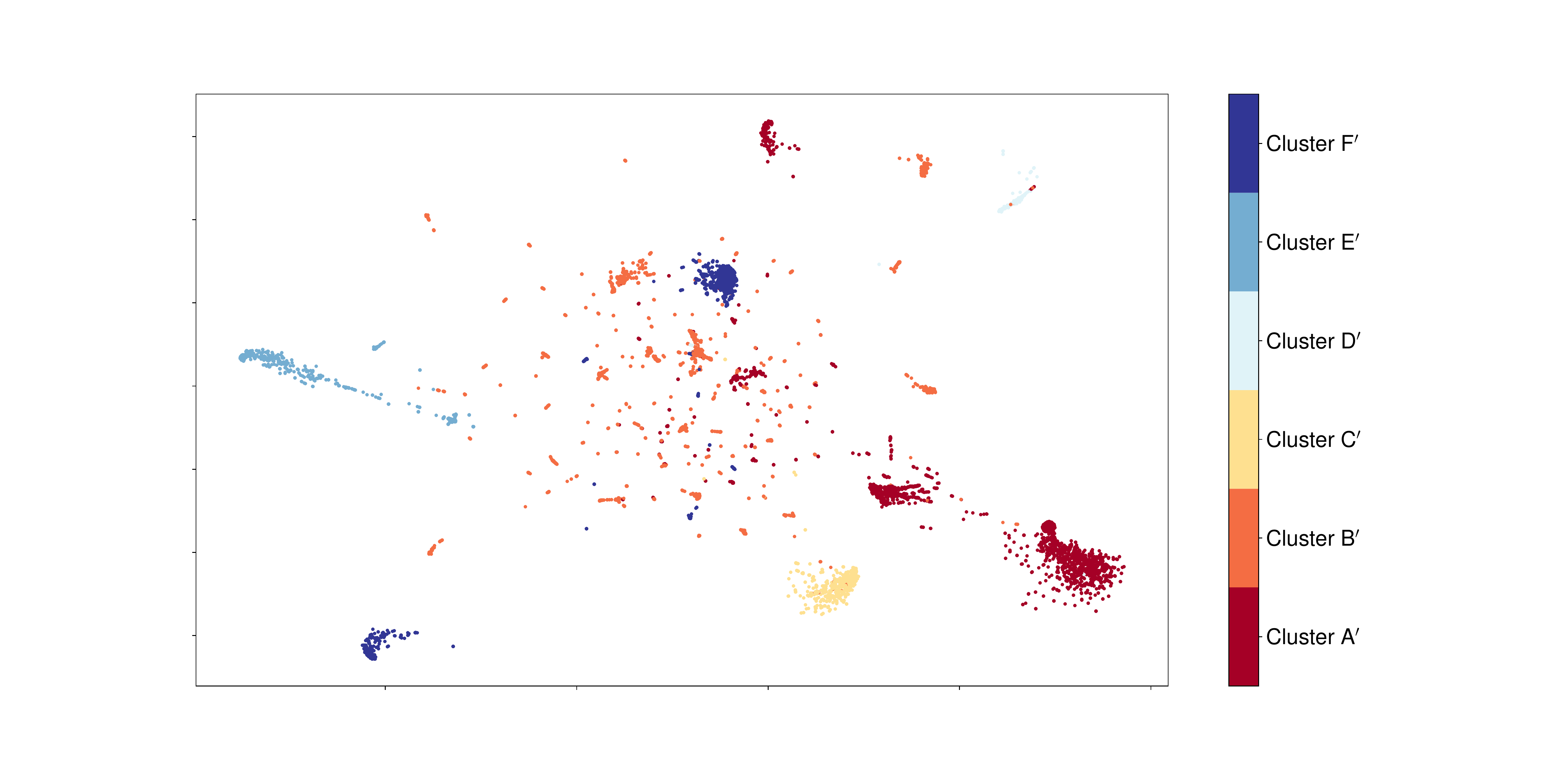}
    \caption{  The 2D UMAP visualization of the US SARS-CoV-2 mutation dataset with 6 distinct clusters. Red, orange, yellow, light blue, blue, and dark blue represent for Cluster A', B', C', D', E', and F', respectively.}  
    \label{fig:1030_US_UMAP}
\end{figure}

 Notably, in \autoref{tab: 1030US co}, Clusters A, B, and C have the same high-frequency co-mutations, indicating that these three clusters may share the same ``root" in the 200D space. However, it is impossible to show the distribution of each cluster in the 200D space. Considering the stability and reliability of UMAP at various reduction ratios, we  employ UMAP to the original US dataset with reduced dimension  2, aiming to observe the distribution of the dataset in the 2D space.  \autoref{fig:1030_US_UMAP} illustrates the 2D visualization of the US dataset with 6 distinct clusters. We can see that there are 3 clusters (Clusters A', B', and F') share the same ``root" located in the middle of the figure, while the other 3 clusters (Clusters C', D', and E') are not. This confirms our deduction about why Clusters A, B, and C have the same high-frequency co-mutations in \autoref{tab: 1030US co}.

\section{Discussion}

 In this section, we compared our past results \cite{wang2020decoding} with our new method to gain a different perspective in clustering with the SNP profiles of COVID-19. In our previous work, a total of 8309 unique single mutations are detected in 15140 SARS-CoV-2 isolates. Here, we also calculate the pairwise distance among 15140 SNP profiles and set the number of clusters to be six. \autoref{tab:world0601} shows the cluster distribution of samples from the 15 countries \cite{wang2020decoding}. The listed countries are the United States (US), Canada (CA), Australia (AU), United Kingdom (UK), Germany (DE), France (FR), Italy (IT), Russia (RU), China (CN), Japan (JP), Korean (KR), India (IN), Spain (ES), Saudi Arabia (SA), and Turkey (TR), and we use Cluster I, II, III, IV, V, and VI to represent six clusters without applying any dimensional reduction algorithm.   \autoref{tab:world0601pca} lists the cluster distribution of samples from the same 15 countries, where we use  I$_p$, II$_p$, III$_p$, IV$_p$, V$_p$, and VI$_p$ to represent six clusters performed by PCA with the reduction ratio to be 1/160. \autoref{tab:world0601umap} lists the cluster distribution of samples from the same 15 countries, where we use  I$_u$, II$_u$, III$_u$, IV$_u$, V$_u$, and VI$_u$ to represent six clusters performed by UMAP with the reduction ratio setting to be 1/160. Noticeably, the SNP profile is focused in Cluster I$_u$, whereas in the non-reduced version, the samples are more spread out. This may be caused by the large number of features, making computed distance between the centroid and each data too similar, and leading to samples being placed in incorrect clusters.

Not surprisingly, PCA and the original method for \cite{wang2020decoding} has nearly identical result. It has been shown in \cite{wang2020decoding} that PCA
is the continuous solution of the cluster indicators in the $K$-means clustering method. On the other hand, UMAP shows a slightly different result. In the PCA method, the distribution is more spread out. In addition, the top occurrence for each country is higher for UMAP. On the other hand, we see that there are more samples in Cluster I$_u$ for UMAP, which may indicate that mutations in Cluster I$_u$ are the main strand.

\begin{figure}[ht]
    \centering
    \includegraphics[width = 1\textwidth]{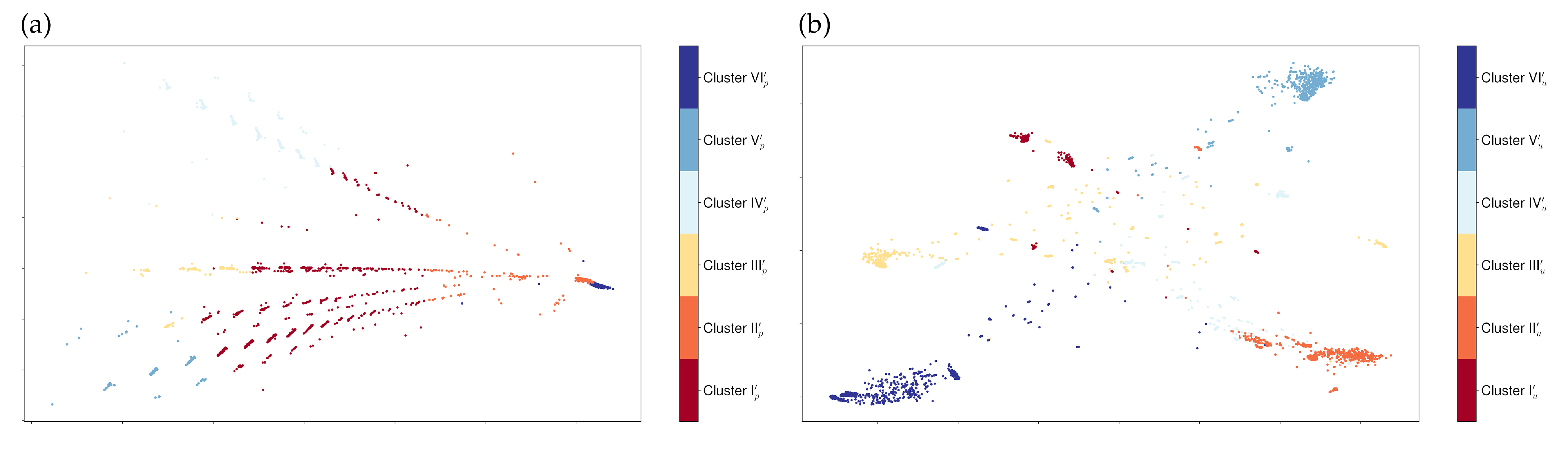}
    \caption{  2D visualizations of the US SARS-CoV-2 mutation dataset up to June 01, 2020 with 6 distinct clusters by applying two different dimensional reduction algorithms. (a) 2D PCA visualization. Red, orange, yellow, light blue, blue, and dark blue represent for Cluster I$_p^{\prime}$, II$_p^{\prime}$, III$_p^{\prime}$, IV$_p^{\prime}$, V$_p^{\prime}$, and VI$_p^{\prime}$, respectively. (b) 2D UMAP visualization. Red, orange, yellow, light blue, blue, and dark blue represent for Cluster I$_u^{\prime}$, II$_u^{\prime}$, III$_u^{\prime}$, IV$_u^{\prime}$, V$_u^{\prime}$, and VI$_u^{\prime}$, respectively.}  
    \label{fig:0601_PCA_UMAP}
\end{figure}

  Moreover,  \autoref{fig:0601_PCA_UMAP} illustrates the 2D visualizations of the US dataset up to June 01, 2020, with 6 distinct clusters by applying two different dimensional reduction algorithms. We can see that the data distribute disorderly under both PCA- and UMAP-assisted $K$-means clustering algorithms.  Specifically, the PCA-assisted algorithm has a really poor clustering performance, while the UMAP-assisted algorithm forms more clear and better clusters than the PCA-assisted algorithm, which is consistent with our previous analysis in Section \ref{sec:Validation}.

% We found that there are 6 clusters of SARS-CoV-2, which is consistent with other related works. In addition, we offer a new perspective by utilizing nonlinear dimensional reduction to obtain a different perspective on the clustering result.

\section{Conclusion} 
The rapid global spread of coronavirus disease 2019 (COVID-19) caused by severe acute respiratory syndrome coronavirus 2 (SARS-CoV-2) has led to genetic mutation stimulated by genetic evolution and adaptation. Up to October 30,  89627 complete SARS-CoV-2 sequences, and a total of 23763 unique SNPs have been detected. Our previous work traced the COVID-19 transmission pathways and analyzed the distribution of the subtypes of SARS-CoV-2 across the world based on 15,140 complete SARS-CoV-2 sequences.  The $K$-means clustering separated the sequences into six distinguished clusters. However, considering the tremendous increase in the number of available SARS-CoV-2 sequences,  an efficient and reliable dimensional reduction method is urgently required.  Therefore, the objective of the present work is to explore the best suited dimension reduction algorithm based on their performance and effectiveness.  Here, a linear algorithm PCA and two non-linear algorithms,  t-distributed stochastic neighbor embedding (t-SNE) and uniform manifold approximation and projection (UMAP), have been discussed.  To evaluate the performance of dimension reduction techniques in clustering, which is an unsupervised problem, we first cast classification problems into clustering problems with labels.  Next, by setting different reduction ratios, we test the effectiveness and accuracy of PCA, t-SNE, and UMAP for $k$-NN and $K$-means using four benchmark datasets.  The results show that overall, UMAP outperforms other two algorithms. 
The major strengths of UMAP is that UMAP-assisted $k$-NN classification and UMAP-assisted $K$-means clustering at various dimension reduction ratios have a consistent performance in terms of accuracy, which proves that UMAP is a stable and reliable dimension reduction algorithm. Moreover, compared to the $K$-means clustering accuracy that does not involve any dimensional reduction, UMAP-assisted $K$-means clustering can improve the accuracy for most cases. Furthermore, when the dimension is reduced to two, the UMAP clustering visualization is clear and elegant. Additionally, UMAP is a relatively efficient algorithm compared to t-SNE. Although PCA is a faster algorithm, its major limitation is its poor performance in accuracy. To be noted, UMAP performs better than PCA and t-SNE for the dataset with a large number of samples, indicating it is the best suited dimensional reduction algorithm for our SARS-CoV-2 mutation dataset. Moreover, we apply the UMAP-assisted $K$-means clustering to the world SARS-CoV-2 mutation dataset (up to October 30, 2020), which displays six distinct clusters. Correspondingly, the same approaches are also applied to the United States SARS-CoV-2 mutation dataset (up to October 30, 2020), resulting in six different clusters as well. Furthermore, we provide a new perspective by utilizing UMAP-assisted $K$-means clustering to analyze our previous SARS-CoV-2 mutation datasets, and the 2D visualization of UMAP-assisted $K$-means clustering of our previous world SARS-CoV-2 mutation dataset (up to June 01, 2020) forms more clear clusters than the PCA-assisted $K$-means clustering. Finally, one of our four datasets was generated by the Jaccard distance representation, which improves both $k$NN classification and $k$-means clustering accuracies on the original dataset.

\section*{Acknowledgment}
This work was supported in part by NIH grant GM126189, NSF Grants DMS-1721024, DMS-1761320, and IIS1900473, Michigan Economic Development Corporation, Bristol-Myers Squibb, and Pfizer. The authors
thank The IBM TJ Watson Research Center, The COVID-19 High Performance Computing Consortium, and
NVIDIA for computational assistance.

\clearpage
\bibliographystyle{abbrv}
%% \bibliographystyle{custom}
%\bibliography{refs}
%\end{document}
\bibliography{refs}

\clearpage
\section{Appendix}

\begin{table}[ht]
	\centering
	\caption{Clusters distribution of the top 25 single mutations of SARS-CoV-2 in the world, collected up to October 30, 2020.}
	\begin{tabular}{cccccccc} \hline \label{1030SNP}
		Top & Position   & Cluster 1 & Cluster 2 & Cluster 3 & Cluster 4 & Cluster 5 & Cluster 6 \\\hline
		Top 1   & 23403 & 1676      & 9301      & 13355     & 17711     & 36351     & 1753      \\
		Top 2   & 14408 & 1676      & 9265      & 13392     & 17728     & 36306     & 1753      \\
		Top 3   & 3037  & 1676      & 9264      & 13350     & 17731     & 36342     & 1753      \\
		Top 4   & 241   & 1669      & 9065      & 13176     & 17549     & 36128     & 1380      \\
		Top 5   & 28881 & 1676      & 64        & 28        & 16509     & 14771     & 1753      \\
		Top 6   & 28882 & 1676      & 15        & 16        & 16508     & 14735     & 1753      \\
		Top 7   & 28883 & 1676      & 13        & 4         & 16513     & 14737     & 1753      \\
		Top 8   & 25563 & 0         & 24        & 13357     & 1022      & 7702      & 0         \\
		Top 9   & 1059  & 0         & 23        & 10919     & 72        & 4797      & 0         \\
		Top 10  & 22227 & 1         & 5         & 25        & 71        & 9211      & 0         \\
		Top 11  & 26801 & 6         & 15        & 14        & 97        & 9094      & 17        \\
		Top 12  & 21255 & 1         & 4         & 8         & 119       & 9101      & 0         \\
		Top 13  & 6286  & 0         & 20        & 5         & 66        & 9092      & 1         \\
		Top 14  & 29645 & 0         & 2         & 9         & 74        & 9069      & 0         \\
		Top 15  & 445   & 0         & 1         & 0         & 53        & 9057      & 0         \\
		Top 16  & 28932 & 0         & 16        & 10        & 52        & 9031      & 1         \\
		Top 17  & 1163  & 1676      & 4         & 0         & 271       & 4566      & 1753      \\
		Top 18  & 22992 & 1675      & 27        & 11        & 930       & 3292      & 1753      \\
		Top 19  & 11083 & 2         & 179       & 415       & 462       & 6046      & 32        \\
		Top 20  & 18555 & 1675      & 13        & 37        & 15        & 3216      & 1752      \\
		Top 21  & 16647 & 1676      & 5         & 9         & 3         & 3200      & 1753      \\
		Top 22  & 23401 & 1675      & 4         & 3         & 17        & 3192      & 1753      \\
		Top 23  & 7540  & 1676      & 0         & 0         & 3         & 3170      & 1753      \\
		Top 24  & 27944 & 0         & 1         & 5         & 49        & 6308      & 0         \\
		Top 25  & 204   & 6         & 6         & 11        & 15        & 5454      & 0        \\\hline
	\end{tabular}
	\label{tab: 1030 mut}
\end{table}

\begin{table}[ht]
	\centering
	\caption{Cluster distributions of SARS-CoV-2 sequences from top 25 countries with the highest number of sequences as of October 30, 2020. The top 25 countries are the United Kingdom (UK), the United States (US), Australia (AU), India (IN), Switzerland (CH), Netherlands (NL), Canada (CA), France (FR), Belgium (BE), Singapore (SG), Spain (ES), Russia (RU), Portugal (PT), Denmark (DK), Sweden (SE), Austria (AT), Japan (JP), South Africa (ZA), Iceland (IS), Brazil (BR), Saudi Arabia (SA), Norway (NO), China (CN), Italy (IT), and Korea (KR).}
	\begin{tabular}{cccccccc}\hline 
		Top   & Country & Cluster 1 & Cluster 2 & Cluster 3 & Cluster 4 & Cluster 5 & Cluster 6 \\\hline
		Top 1  & UK      & 16        & 2926      & 1112      & 9530      & 22959     & 48        \\
		Top 2  & US      & 1         & 1798      & 8620      & 2209      & 9760      & 0         \\
		Top 3  & AU      & 1652      & 247       & 346       & 286       & 4328      & 1700      \\
		Top 4  & IN      & 0         & 415       & 138       & 683       & 708       & 0         \\
		Top 5  & CH      & 0         & 580       & 114       & 411       & 827       & 0         \\
		Top 6  & NL      & 0         & 397       & 80        & 283       & 942       & 0         \\
		Top 7  & CA      & 1         & 193       & 447       & 166       & 572       & 2         \\
		Top 8  & FR      & 5         & 220       & 451       & 176       & 223       & 2         \\
		Top 9  & BE      & 0         & 285       & 66        & 269       & 274       & 0         \\
		Top 10 & SG      & 0         & 35        & 50        & 115       & 675       & 0         \\
		Top 11 & ES      & 0         & 406       & 14        & 81        & 365       & 0         \\
		Top 12 & RU      & 0         & 62        & 29        & 539       & 101       & 0         \\
		Top 13 & PT      & 0         & 114       & 31        & 334       & 166       & 0         \\
		Top 14 & DK      & 0         & 96        & 359       & 32        & 89        & 0         \\
		Top 15 & SE      & 0         & 84        & 182       & 219       & 79        & 0         \\
		Top 16 & AT      & 0         & 142       & 108       & 180       & 129       & 0         \\
		Top 17 & JP      & 0         & 25        & 26        & 236       & 242       & 0         \\
		Top 18 & ZA      & 0         & 72        & 11        & 299       & 134       & 0         \\
		Top 19 & IS      & 0         & 137       & 67        & 54        & 162       & 0         \\
		Top 20 & BR      & 0         & 30        & 14        & 154       & 194       & 0         \\
		Top 21 & SA      & 0         & 5         & 193       & 81        & 98        & 0         \\
		Top 22 & NO      & 0         & 42        & 23        & 61        & 224       & 0         \\
		Top 23 & CN      & 0         & 18        & 8         & 43        & 258       & 0         \\
		Top 24 & IT      & 0         & 120       & 11        & 129       & 66        & 0         \\
		Top 25 & KR      & 0         & 10        & 21        & 26        & 260       & 0        \\\hline
	\end{tabular}
	\label{tab: 1030Country}
\end{table}

\begin{table}[ht]
	\centering
	\caption{Clusters distribution of the top 25 single mutations of SARS-CoV-2 in the United States, collected up to October 30, 2020.}
	\begin{tabular}{cccccccc}  \hline 
		Top & Position   & Cluster A & Cluster B & Cluster C & Cluster D & Cluster E & Cluster F \\\hline 
		Top 1   & 23403 & 6686      & 8110      & 1418      & 486       & 3         & 2757      \\
		Top 2   & 14408 & 6688      & 8100      & 1418      & 486       & 3         & 2752      \\
		Top 3   & 3037  & 6674      & 8097      & 1417      & 486       & 2         & 2756      \\
		Top 4   & 241   & 6562      & 8033      & 1403      & 470       & 2         & 2695      \\
		Top 5   & 25563 & 4906      & 6405      & 1417      & 485       & 0         & 8         \\
		Top 6   & 1059  & 4235      & 5831      & 1417      & 485       & 1         & 3         \\
		Top 7   & 27964 & 4         & 1880      & 1402      & 0         & 1         & 1         \\
		Top 8   & 28881 & 16        & 408       & 2         & 2         & 1         & 2733      \\
		Top 9   & 28882 & 9         & 401       & 1         & 1         & 0         & 2732      \\
		Top 10  & 28883 & 0         & 399       & 0         & 1         & 0         & 2738      \\
		Top 11  & 28144 & 8         & 728       & 0         & 0         & 1486      & 2         \\
		Top 12  & 8782  & 1         & 728       & 0         & 0         & 1482      & 1         \\
		Top 13  & 18060 & 14        & 299       & 0         & 0         & 1469      & 8         \\
		Top 14  & 17858 & 3         & 296       & 0         & 0         & 1460      & 2         \\
		Top 15  & 17747 & 1         & 308       & 0         & 0         & 1420      & 3         \\
		Top 16  & 20268 & 1125      & 389       & 0         & 0         & 0         & 0         \\
		Top 17  & 10319 & 0         & 1104      & 381       & 0         & 0         & 0         \\
		Top 18  & 28854 & 897       & 528       & 17        & 0         & 0         & 0         \\
		Top 19  & 19839 & 0         & 199       & 0         & 1         & 0         & 1058      \\
		Top 20  & 29870 & 237       & 776       & 36        & 9         & 55        & 113       \\
		Top 21  & 36    & 32        & 694       & 22        & 117       & 192       & 105       \\
		Top 22  & 29784 & 4         & 157       & 0         & 0         & 1         & 794       \\
		Top 23  & 15933 & 0         & 153       & 0         & 0         & 0         & 705       \\
		Top 24  & 11083 & 142       & 590       & 32        & 12        & 21        & 55        \\
		Top 25  & 11916 & 1         & 822       & 9         & 0         & 0         & 12       \\\hline 
	\end{tabular}
	\label{tab: 1030US mut}
\end{table}

\begin{table}[ht] 
	\centering 
	\caption{Cluster statistics for states with more than 50 SARS-CoV-2 genome samples.}
	\begin{tabular}{ccccccc} \hline 
		State                & Cluster A & Cluster B & Cluster C & Cluster D & Cluster E & Cluster F \\\hline 
		New Hampshire        & 330       & 1743      & 25        & 442       & 870       & 1082      \\
		New York             & 1461      & 918       & 108       & 2         & 210       & 451       \\
		Kansas               & 425       & 1461      & 302       & 0         & 3         & 332       \\
		Wisconsin            & 328       & 1189      & 20        & 0         & 12        & 192       \\
		Missouri             & 944       & 592       & 16        & 2         & 28        & 73        \\
		North Dakota         & 310       & 701       & 183       & 0         & 121       & 41        \\
		Alaska               & 55        & 505       & 319       & 0         & 1         & 7         \\
		Arizona              & 442       & 267       & 28        & 1         & 16        & 104       \\
		South Dakota         & 260       & 385       & 56        & 0         & 14        & 140       \\
		Utah                 & 266       & 236       & 100       & 0         & 46        & 30        \\
		District of Columbia & 232       & 222       & 7         & 5         & 15        & 53        \\
		Hawaii               & 79        & 148       & 3         & 0         & 1         & 16        \\
		Kentucky             & 91        & 89        & 36        & 0         & 2         & 11        \\
		Maine                & 59        & 85        & 0         & 11        & 4         & 17        \\
		Delaware             & 44        & 87        & 10        & 0         & 17        & 10        \\
		Wyoming              & 12        & 47        & 5         & 11        & 41        & 41        \\
		Arkansas             & 24        & 49        & 22        & 0         & 0         & 46        \\
		Oregon               & 41        & 69        & 6         & 0         & 10        & 12        \\
		South Carolina       & 10        & 33        & 71        & 4         & 7         & 0         \\
		Texas                & 35        & 71        & 0         & 1         & 12        & 2         \\
		Tennessee            & 45        & 27        & 26        & 1         & 14        & 7         \\
		Washington           & 18        & 72        & 4         & 1         & 5         & 6         \\
		Minnesota            & 69        & 17        & 0         & 0         & 7         & 4         \\
		New Mexico           & 40        & 28        & 10        & 0         & 0         & 11        \\
		Mississippi          & 53        & 28        & 0         & 0         & 1         & 1         \\
		Vermont              & 22        & 42        & 0         & 1         & 8         & 3         \\
		Indiana              & 6         & 62        & 1         & 0         & 0         & 0         \\
		New Jersey           & 27        & 5         & 5         & 0         & 5         & 9         \\
		California           & 21        & 14        & 0         & 0         & 0         & 15       \\\hline 
	\end{tabular}
	\label{USstats}
\end{table}

\begin{table}[ht]
	\centering
	\caption{The world wide clusters from SARS-CoV-2 genome data available up to June 01, 2020. The listed countries are the United States (US), Canada (CA), Australia (AU), United Kingdom (UK), Germany (DE), France (FR), Italy (IT), Russia (RU), China (CN), Japan (JP), Korean (KR), India (IN), Spain (ES), Saudi Arabia (SA), and Turkey (TR). \cite{wang2020decoding}.}
	\begin{tabular}{ccccccc} \hline
		Country	& Cluster I	& Cluster II & Cluster III & Cluster IV & Cluster V & Cluster VI \\\hline
		US & 844 & 311 & 488 & 156 & 1813 & 975 \\
		CA & 12  & 29  & 17 & 16 & 19 & 41 \\
		AU & 163 & 149 & 410 & 135 & 146 & 77 \\
		UK & 539 & 875 & 908 & 1532 & 119 & 3 \\
		DE & 10  & 20  & 21 & 38 & 42 & 0 \\
		FR & 41  & 85  & 14 & 12 & 82 & 0 \\
		IT & 26  & 24  & 9 & 17 & 0 & 0 \\
		RU & 10  & 27  & 1 & 109 & 3 & 0 \\
		CN & 8   & 3   & 215 & 1 & 1 & 25 \\
		JP & 0   & 3   & 68 & 20 & 3 & 0 \\
		KR & 0   & 0   & 28	0 & 0 & 0 & 0 \\
		IN & 93  & 69  & 141 & 10 & 3 & 0 \\
		ES & 27  & 100 & 74 & 25 & 3 & 2 \\
		SA & 14  & 31  & 9 & 1 & 2 & 0 \\
		TR & 25  & 3   & 24 & 9 & 0 & 0 \\\hline
	\end{tabular}
	\label{tab:world0601}
\end{table}
\begin{table}[ht]
	\centering
		\caption{The world wide clusters from SARS-CoV-2 genome data available up to June 01, 2020 using PCA embedding with reduction ratio of 1/160.}
	\begin{tabular}{ccccccc}\hline
		Country	& Cluster I$_p$	& Cluster II$_p$ & Cluster III$_p$ & Cluster IV$_p$ & Cluster V$_p$ & Cluster VI$_p$ \\ \hline
		US & 915 & 489 & 239 & 156  & 1813 & 975 \\
		CA & 14  & 17  & 27  & 16   & 19   & 41  \\
		AU & 164 & 414 & 143 & 136  & 146  & 77  \\
		UK & 543 & 908 & 857 & 1546 & 119  & 3   \\
		DE & 10  & 21  & 20  & 38   & 42   & 0   \\
		FR & 46  & 14  & 80  & 12   & 82   & 0   \\
		IT & 26  & 9   & 24  & 17   & 0    & 0   \\
		RU & 10  & 1   & 27  & 109  & 3    & 0   \\
		CN & 8   & 213 & 3   & 1    & 1    & 24  \\
		JP & 0   & 68  & 3   & 20   & 3    & 0   \\
		KR & 0   & 28  & 0   & 0    & 0    & 0   \\
		IN & 95  & 141 & 67  & 10   & 3    & 0   \\
		ES & 27  & 74  & 100 & 25   & 3    & 2   \\
		SA & 30  & 9   & 15  & 1    & 2    & 0   \\
		TR & 27  & 24  & 1   & 9    & 0    & 0   \\\hline
	\end{tabular}
	\label{tab:world0601pca}
\end{table}
\begin{table}[ht]
	\centering
	\caption{The world wide clusters from SARS-CoV-2 genome data available up to June 01, 2020 using UMAP embedding with reduction ratio of 1/160}
	\begin{tabular}{c c c c c c c} \hline
		Country	& Cluster I$_u$	& Cluster II$_u$ & Cluster III$_u$ & Cluster IV$_u$ & Cluster V$_u$ & Cluster VI$_u$ \\ \hline
		US & 2446 & 1096 & 90 & 751 & 110 & 94 \\
		CA & 71 & 15 & 9 & 35 & 1 & 3 \\
		AU & 784 & 94 & 64 & 18 & 83 & 37 \\
		UK & 2171 & 115 & 828 & 2 & 534 & 326 \\
		DE & 57  & 40 & 14 & 0 & 5  & 15 \\
		FR & 163 & 45 & 10 & 0 & 11 & 5  \\
		IT & 13  & 1  & 35 & 0 & 5  & 22 \\
		RU & 92  & 2  & 49 & 0  & 0  & 7 \\
		CN & 178 & 28 & 6  & 10 & 22 & 6 \\
		JP & 36  & 0  & 11 & 0  & 47 & 0\\
		KR & 18  & 0 & 0  & 1 & 9 & 0  \\
		IN & 232 & 3 & 7  & 0 & 2 & 72 \\
		ES & 205 & 2 & 12 & 0 & 7 & 5 \\
		SA & 56  & 0 & 1  & 0 & 0 & 0 \\
		TR & 56  & 1 & 4  & 0 & 0 & 0 \\ \hline
	\end{tabular}
	\label{tab:world0601umap}
\end{table}

% \end{multicols}
\end{document}